\documentclass[preprint,10pt]{elsarticle}

\usepackage{amssymb}
\usepackage{amsmath}

\usepackage[left=1in, right=1in, top=1in, bottom=1in]{geometry} 
\usepackage{xcolor} 

\usepackage{array,multirow}  
\DeclareUnicodeCharacter{03F5}{\ensuremath{\epsilon}} 
\usepackage[colorlinks=true,linkcolor=blue,citecolor=blue,urlcolor=blue]{hyperref}  
\usepackage{setspace} 

\journal{Methods}
\begin{document}
\begin{frontmatter}

\title{\textit{In-vitro} measurements coupled with \textit{in-silico} simulations 
for stochastic calibration and uncertainty quantification  
of the mechanical response of biological materials}

\author{Mahmut Pekedis}
\ead{mahmut.pekedis@ege.edu.tr}

\affiliation{organization={Department of Mechanical Engineering, Ege University}, 
            addressline={35040, Bornova}, 
            city={Izmir},
            postcode={35040}, 
            country={Turkey}}

\begin{abstract}

\small This study proposes a simple and practical approach based on \textit{in-vitro} measurements and \textit{in-silico} simulations using the likelihood-free Bayesian inference with the finite element method simultaneously for stochastic calibration and uncertainty quantification of the mechanical response of biological materials. We implement the approach for distal, middle, and proximal human Achilles tendon specimens obtained from diabetic patients post-amputation. A wide range of \textit{in-vitro} loading conditions are considered, including one-step and two-step relaxation, as well as incremental cyclic loading tests. \textit{In-silico} simulations are performed for the tendons assuming a fiber-reinforced viscoelastic response, which is modeled for the ground matrix and fiber components. Initially, the calibration of the specimen-specific parameters is predicted using Bayesian optimization and the sensitivity of each parameter is evaluated using the Sobol index and random forest. Then, these parameters are used as priors, and coupled with \textit{in-vitro} data in simulation-based approximate Bayesian computation to calibrate and quantify the uncertainty parameters for three loading cases. The results demonstrate that \textit{in-silico} simulations using the posterior parameters of approximate Bayesian computation, can capture the uncertainty bounds of \textit{in-vitro} measurements. This approach provides a useful framework for stochastic calibration of constitutive material model parameters without the need to derive a likelihood function, regardless of the specimen’s geometry or loading conditions.
\end{abstract}
\end{frontmatter}


\section{Introduction}

\textit{In-silico} mechanical modeling and simulation of biological materials have the potential to facilitate advancements in every aspect of the medical field. It offers numerous advantages, such as being a promising tool for the pre-clinical testing of medical devices, helping to diagnose and evaluate targeted treatments for patient-specific applications, and supplementing experimental investigations when \textit{in-vivo} experiments are not possible \cite{viceconti_possible_2021, smit_silico_2024, VV40, pekedis_2025_2}. It can also model patient-specific tissue or organ-level behavior and, more generally, account for variations within a population. One such factor that significantly influences the \textit{in-silico} simulation outcome is the constitutive material model, as well as the parameters used in it. The model and its calibrated biomechanical parameters should accurately reflect and mimic the tissue's behavior. These parameters are typically calibrated using a non-linear least squares fit of the model to mechanical testing data from experiments such as tension \cite{ogden_fitting_2004, pekedis_2021}, compression \cite{budday_fifty_2020}, and biaxial testing \cite{niestrawska_microstructure_2016}.  For most calibration procedures, the choice of a constitutive model is assessed by the quality of fit (e.g., \(R^2\) and the value of the root mean square error RMSE). This technique is efficient and powerful for patient or specimen-specific calibration. However, because most biological materials are nonhomogeneous and anisotropic, deploying experimental tests using standard mechanical tests may not be enough for mechanical characterization of the tissue. Therefore, calibration of the model parameters directly from the experimental data may not always be possible or relevant. In some studies, the inverse technique is performed to overcome these issues \cite{MENGONI2021105, narayanan_inverse_2021}. In brief, the inverse approach is considered as the estimation of input parameters (material properties, geometric features, loading and boundary conditions, etc.) from output observations (displacement, strains, velocity). Once a forward model is created, the input parameters are predicted using experimental observations with an optimization technique. This method is powerful for patient/specimen-specific material calibration when inter-variability is not considered.

In addition to inhomogeneous characteristics, biological materials often exhibit intrinsic variability in their biomechanical response when subjected to external loads. The inter-subject or inter-specimen variability and uncertainty in biomechanical properties of these materials are high, which may be influenced by many factors such as the donor's age, genetics, the tissue’s pathological state, gender, and other specific factors related to the tissue characteristics. For \textit{in-vitro} samples, embalming medium, temperature, humidity, and duration are the additional factors that can affect the biomechanical properties. Oftentimes, \textit{in-silico} model is hindered not only by the incompleteness of the theoretical approach but also by uncertainties due to mentioned factors that arise in the experimental observations. Thus, there is an urgent need to ensure that it can capture the intrinsic population of variability and not only be representative of `mean' individuals. 

Uncertainty can also arise due to lack of knowledge of the physical model of interest (insufficient initial and boundary conditions), as an addendum to the inherent variation in material properties \cite{oberkamf, henninger_validation_2010}. However, most of the studies implemented in the field of biomechanics have mainly investigated uncertainties due to variabilities in material properties using the Bayesian framework \cite{mihai_LA, laz}. In this framework, once the prior distribution of the parameters is constructed, in conjunction with a likelihood function, the posterior distribution is determined using random sampling methods. There has been much research in the field of brain mechanics \cite{madireddy_bayesian_2015, teferra_bayesian_2019, brewick_uncertainty_2018, staber_stochastic_2017, mihai_LA}. Furthermore, several studies have established the utility of Bayesian inference in solid mechanics and soft tissues, addressing both parameter estimation and model selection. For instance, Madireddy et al.~\cite{madireddy_bayesian_2015, madireddy_bayesian_2016} presented a Bayesian model selection and calibration framework for hyperelastic soft tissue models. Haughton et al.~\cite{haughton_bayesian_2022} used Bayesian techniques to quantify uncertainty in microstructural tendon models. Oden et al.~\cite{oden_selection_2013} and Chiachío et al.~\cite{chiachio_bayesian_2015} implemented Bayesian methods for selecting and validating phenomenological and damage progression models. Rappel et al.~\cite{rappel_elastoplastic_2019} provided detailed frameworks to identify the parameters of the elastoplastic material under uncertainty. In addition, James Beck and colleagues have applied Bayesian learning to structural health monitoring and inverse problems in mechanics ~\cite{chenyue_robust_2023, huang_multitask_2019,huang_sequential_2021,Xianghao_Adaptive_2025}. Furthermore, Babuska et al.~\cite{babuska_bayesian_2016} presented a Bayesian model comparison approach for metallic fatigue data, while Xue et al.~\cite{xue_damage_2020} proposed sparse Bayesian learning framework for robust damage localization using guided-wave testing. These studies primarily rely on standard Bayesian frameworks that require an explicit likelihood function. However, the likelihood function can be difficult or computationally expensive to derive for nonhomogeneous materials under complex biomechanical loading conditions. To overcome these such issues, an approximate Bayesian computation (ABC) approach that does not use the likelihood function has been proposed \cite{toni_simulation-based_2010, schalte_pyabc_2022} and implemented recently in some fields including structural health monitoring \cite{Guofeng} medical decision making \cite{Huaqing} and operation research management \cite{garud}. However, the coupling of \textit{in-silico} simulations for stochastic calibration and uncertainty quantification using likelihood-free Bayesian computation in the field of mechanics of biological materials with high inter-specimen variability has been largely unaddressed to date, to the best of our knowledge. The most attractive aspect of this approach is that the likelihood function is not required to be explicitly represented in a closed analytical form. Once an \textit{in-silico} model is built to mimic the \textit{in-vitro} conditions, it directly bypasses the need for explicit likelihood evaluation and relies on comparing the simulated data from the \textit{in-silico} model with experimental observations from the \textit{in-vitro} test. This approach can be implemented in a wide range of \textit{in-vivo} and \textit{in-vitro} applications, ranging from predicting the mechanical characteristics change of tissue due to the risk of pathogenic disease to planning surgical procedures. We note that this study is not targeted to develop or present a constitutive material model for connective tissue, but rather to show how likelihood-free inference can be used directly with the numerical simulator and measurements. Any suggested constitutive model can be directly embedded in the simulator.

In this work, we suggest a simple approach based on \textit{in-vitro} measurements, \textit{in-silico} simulations, and approximate Bayesian computation for stochastic calibration and uncertainty quantification of material parameters. The paper is structured as follows. First, a specimen-specific calibration procedure using Bayesian optimization is presented. Next, the sensitivity analysis and stochastic calibration steps implemented in ABC are detailed in the following section. Finally, the results obtained from the methodology are presented and summarized, along with some concluding remarks.

\section{Materials and Methods}
\subsection{\textit{In-vitro} measurements}
\begin{figure*}[!ht]
\centering
\includegraphics [width=6in] {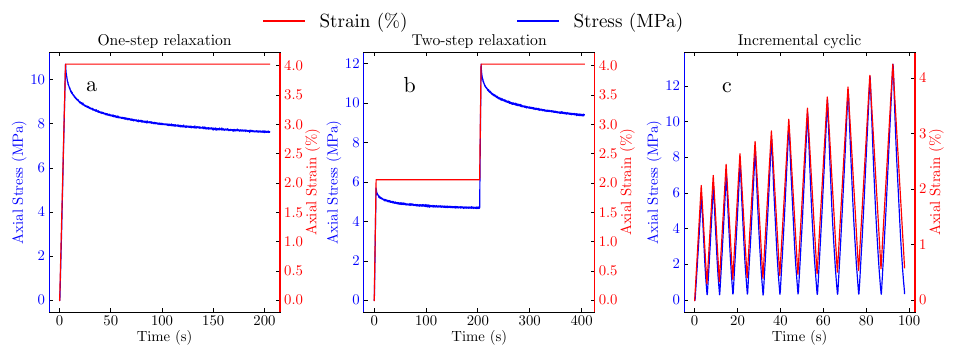}
\caption {Typical experimental \textit{in-vitro} measurements observed for a specimen. (a) One-step relaxation. (b) Two-step relaxation. (c) Incremental cyclic loading}
\label{Figure_1}
\end{figure*}

The approach is performed on recently published data of Achilles tendons  from diabetic patients who underwent below or above knee amputation \cite{pekedis_2025}. In summary, the tendons were harvested  from the proximal end to the insertion of the Achilles tendon at the calcaneus. Their length was approximately 120 to 130 cm and were partitioned into three equal sections (distal, middle, and proximal site). Then, they were resectioned to form a standard dog bone specimen. The dimensions of each specimen at the measurement site had a gauge length of approximately 10 mm, a width of $\sim$ 8 mm and a thickness of $\sim$ 4 mm.  Uniaxial displacement-type loading was performed to collect force measurements. First, samples were preloaded at 0.1 MPa to achieve parallel alignment of the fibrils, and then preconditioned for 10 cycles with a strain amplitude of 2\% strain at a strain rate of 0.5\% $s^{-1}$. The test protocols were as follows:
\begin{itemize}
    \item One-step relaxation: Initiates with a ramp strain of 0.6\% ${s^{-1}}$ followed by a constant strain of 4\% for a relaxation duration of 200 s (Figure \ref{Figure_1}a).
    \item Two-step relaxation: Two consecutive steps of 2\% strain (ramp strain of 0.6\% $s^{-1}$) followed by a 200-s relaxation time (Figure \ref{Figure_1}b).
    \item Incremental cyclic: Incremental cyclic loading ranging from a strain amplitude of 2\% to 4.25\% with an increment of 0. 2\% strain in each cycle (Figure \ref{Figure_1}c). 
\end{itemize}
Further details of the tests and \textit{in-vitro} measurements are reported in previous work \cite{pekedis_2025}. A typical stress-strain observation of a specimen for three loading scenarios is presented in Figure \ref{Figure_1}. The force and displacement signals have a sampling rate of 100 Hz. \textit{In-vitro} measurements include 81 datasets (9 specimens $ \texttt{x}$ 3 sites $\texttt{x}$ 3 test protocols). Each of these data sets consists of three dimensions: the first is time, the second is displacement, and the third is force. The data sets are organized and saved as PYTHON \texttt{pickle} for further serialization processes between \textit{in-vitro} measurements and \textit{in-silico} simulations for sensitivity, optimization and Bayesian inference tasks.
\subsection{Fiber reinforced visco elastic model}
Soft biological tissues, such as the Achilles tendon, are collagenous tissues that connect the calf muscle to the calcaneus. Their primary function is to transmit tensile forces and act as mechanical dampers. These tissues are primarily composed of water, Type I collagen fibers, and proteoglycans. They can be viewed as fibrous composites, consisting of a ground matrix embedded with families of collagen fibers.  The tendon is modeled using the fiber-reinforced viscoelastic (FRVE) approach, in which the solid matrix is divided into a nonfibrillar part representing the proteoglycan matrix and a fibrillar part representing the collagen fibers. The biomechanical response of these constituents, which drives the approximate Bayesian computation (ABC), is simulated using the FEBio framework \cite{maas_febio_2012}.  In this FRVE model, two solid mixtures are considered: the ground matrix, representing the proteoglycan matrix, and the collagen fibrils, representing the fibers. Viscoelastic effects are incorporated for both constituents. Specifically, the ground matrix is modeled as a compressible  visco linear elastic material, whereas the collagen fibrils are assumed to be compressible  visco nonlinear elastic materials that respond only to tensile loads.
The relaxation function for the fibril and matrix components is given by:
\begin{equation}
\label{eq:prony}
    G(t) = g_0 + \sum_{i = 1}^n {g_i e^{-\frac{t}{\tau_i}} } 
\end{equation}
where $t$ is time, $g_i$ and $\tau_i$ are $i$'th viscoelastic characteristic coefficients. The strain energy density function for the fiber component is defined as:
\begin{equation}
    \psi_n =
\begin{cases}
0, & I_n < 1 \quad \text{(under compression)} \\
\frac{\xi}{\alpha \beta} \left(\exp\left[\alpha(I_n - 1)^{\beta}\right] - 1\right), & 1 \le I_n \le I_0 \quad \text{(nonlinear toe region)} \\
B(I_n - I_0) - E_f \left(I_n^{1/2} - I_0^{1/2}\right) + \frac{\xi}{\alpha \beta} \left(\exp\left[\alpha(I_0 - 1)^{\beta}\right] - 1\right), & I_n > I_0 \quad \text{(linear region)}
\end{cases}
\end{equation}
where $E_f$ is the fiber modulus in the linear region, $\alpha$ is the coefficient of the exponential term, and $\beta$ is the power-law exponent that controls the nonlinearity in the toe region. The scalar quantity $\lambda_n$ represents the fiber stretch ratio, defined as the ratio of the current fiber length to its reference length. The related invariant $I_n$ corresponds to the square of the fiber stretch ratio, $I_n = \lambda_n^2$. A threshold value $I_0 = \lambda_0^2$ defines the transition from the nonlinear toe region to the linear region of the material response. The coefficients $\xi$ and $B$ are defined to ensure continuity of the strain energy function:
\begin{equation}
    \xi = \frac{E_f(I_0 - 1)^{2 - \beta} \exp\left[-\alpha(I_0 - 1)^{\beta}\right]}{4 I_0^{3/2} \left( \beta - 1 + \alpha \beta (I_0 - 1)^{\beta} \right)},
\end{equation}
\begin{equation}
    B = E_f  \frac{2 I_0 \left( \beta - \frac{1}{2} + \alpha \beta (I_0 - 1)^{\beta} \right) - 1}{4 I_0^{3/2} \left( \beta - 1 + \alpha \beta (I_0 - 1)^{\beta} \right)}.
\end{equation}
\textit{In-vitro} measurements showed that $\lambda_0$ is approximately 1.01 and assumed to be a fixed parameter. For this material type, the fiber modulus at the strain origin reduces to zero unless \(\beta = 2\). Thus, we set \(\beta = 2\) in the \textit{in-silico} simulation. In summary, the matrix is assumed to exhibit viscoelastic behavior, while the fiber is modeled as a viscoelastic component that provides tensile reinforcement, using the \texttt{fiber-exp-pow-linear} material model available in FEBio. This approach captures both the time-dependent behavior and the load-bearing property of the fibers within the composite tissue. For additional theoretical background, see \cite{maas_febio_2012, weiss_1996}. The model parameters, their Bayesian optimization search ranges, and prior distributions are summarized in Table~\ref{table_1}.
\begin{table}[ht]
\centering
\small
\caption{Parameter space used in Bayesian optimization (BO) for the three sites.\label{table_1}}
\label{tab:parameters}
\renewcommand{\arraystretch}{1.3}
\setlength{\tabcolsep}{8pt}
\begin{tabular}{lllc}
\hline
Constituent & Parameter & Description & Ranges (Min, Max) \\
\hline
\multirow{8}{*}{Fiber} 
 & \textit{E}\textsubscript{f} (MPa) & Fiber modulus & (160, 275) \\
 & \textit{a} (-) & Exponential argument & (0, 10E3)\\
 & \textit{g}\textsubscript{1f} (MPa) & 1\textsuperscript{st} viscoelastic characteristic & (0.8, 1) \\
 & \textit{g}\textsubscript{2f} (MPa) & 2\textsuperscript{nd} viscoelastic characteristic & (0.8, 2.1) \\
 & \textit{g}\textsubscript{3f} (MPa) & 3\textsuperscript{rd} viscoelastic characteristic & (0.8, 2.1) \\
 & $\tau_{1\text{f}}$ (s) & 1\textsuperscript{st} viscoelastic relaxation & (0.1, 2) \\
 & $\tau_{2\text{f}}$ (s) & 2\textsuperscript{nd} viscoelastic relaxation & (2, 20) \\
 & $\tau_{3\text{f}}$ (s) & 3\textsuperscript{rd} viscoelastic relaxation  & (20, 200) \\
\hline
\multirow{8}{*}{Ground Matrix} 
 & \textit{E}\textsubscript{m} (MPa) & Elastic modulus & (11, 70) \\
 & \textit{g}\textsubscript{1m} (MPa) & 1\textsuperscript{st} viscoelastic characteristic  & (0.8, 1.5) \\
 & \textit{g}\textsubscript{2m} (MPa) & 2\textsuperscript{nd} viscoelastic characteristic  & (0.8, 2.1) \\
 & \textit{g}\textsubscript{3m} (MPa) & 3\textsuperscript{rd} viscoelastic characteristic & (0.8, 2.1) \\
 & $\tau_{1\text{m}}$ (s) & 1\textsuperscript{st} viscoelastic relaxation & (0.1, 2) \\
 & $\tau_{2\text{m}}$ (s) & 2\textsuperscript{nd} viscoelastic relaxation & (3, 20) \\
 & $\tau_{3\text{m}}$ (s) & 3\textsuperscript{rd} viscoelastic relaxation & (20, 200) \\
 & $v_{\text{m}}$ (-) & Poisson’s ratio & (0, 0.4) \\
\hline
\multicolumn{4}{l}{Min and Max values represent the search ranges used in BO.}
\end{tabular}
\end{table}


\subsection{\textit{In-silico} simulations}
\begin{figure*}[!ht]
\centering
\includegraphics [width=6.5in] {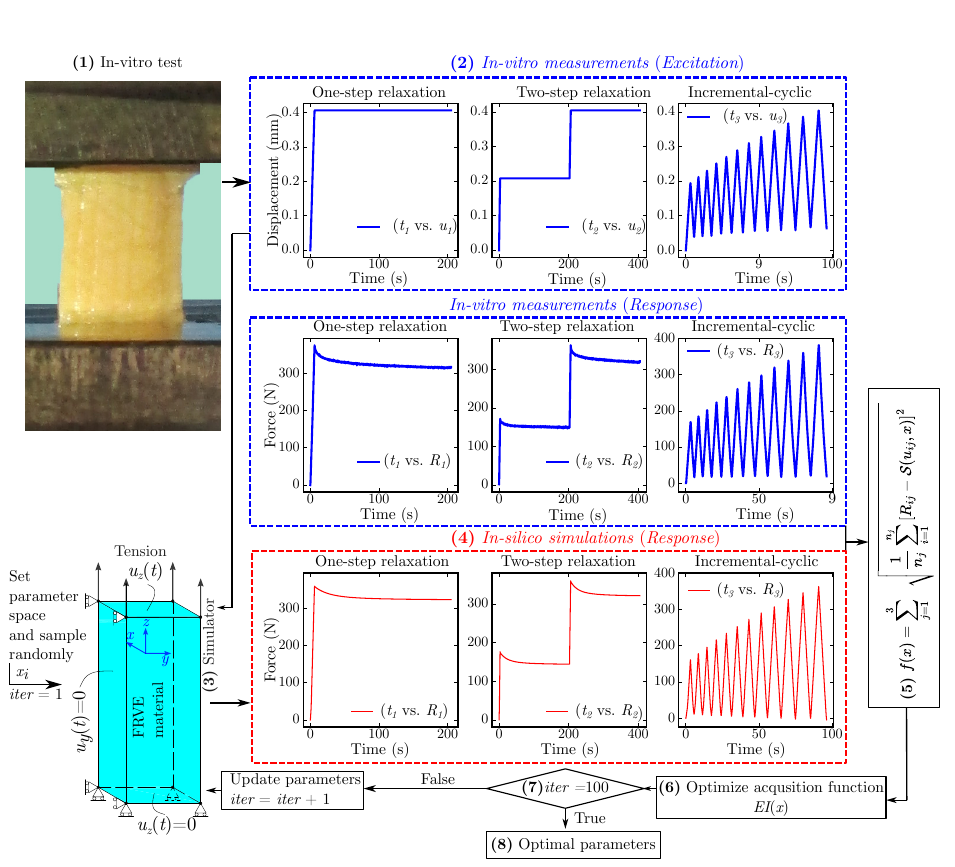}
\caption {Specimen-specific parameter calibration steps using \textit{in-vitro} measurements, \textit{in-silico} simulations, and Bayesian optimization}
\label{Figure_2}
\end{figure*}

Numerical methods such as the finite element method, meshless methods, or the finite volume method can be used as simulators to bridge the gap between \textit{in-vitro} measurements and Bayesian computation for stochastic calibration and quantification of uncertainty in the biomechanical response of biological materials. Here, we used the finite element method as a simulator. The geometric dimensions of the \textit{in-vitro} sample are approximately $10 \times 8 \times 4 \, \text{mm}$ (gauge length $\times$ width $\times$ thickness). To reduce computational cost, the sample is modeled using a single 3D hexahedral element, with symmetry imposed in the \textit{y}-direction by constraining the left face ($u_y(t) = 0$), as shown in Figure~\ref{Figure_2}. The bottom nodes are constrained vertically ($u_z(t) = 0$), and the top nodes are displacement-controlled using measured displacements $u_z(t)$, consistent with the \textit{in-vitro} test protocols. We note that additional \textit{in-silico} outputs, such as the deformation gradient, the strain distribution, or the direct stress, can also be extracted and used with ABC if the corresponding measurements are available. Typically, full-field deformation measurements can be collected via image-based techniques, such as digital image correlation. We believe that higher-dimensional data from \textit{in-vitro} observations will lead to more sensitive calibrated parameters.
\subsection{Bayesian optimization and sensitivity analysis}
To determine specimen-specific biomechanical parameters, Bayesian optimization (BO) is implemented. BO was selected primarily due to the high computational cost of finite element simulations involved in evaluating the objective function \( f(x) \). Each evaluation requires solving a full-scale FE model, making traditional methods such as grid search or evolutionary algorithms computationally impractical. Beyond efficient parameter calibration, BO also serves to inform the parameter set values used as priors in the ABC-based uncertainty quantification framework. It is an iterative framework focused on optimizing an expensive and unknown objective function \(f(x)\), where \( x \in \mathbb{R}^d \) is a vector of \( d \) parameters. The goal is to solve
\begin{equation}
\mathbf{x}^{*} = \arg\min_{\mathbf{x} \in \mathcal{A}}f(x)
\end{equation}
where \(\mathcal{A}\) is the search space. Gradient information is not available for \(f(x)\) and is considered a ``black box" function. The core of BO  is based on the construction of a probabilistic surrogate model, a Gaussian process (GP), to approximate \(f(x)\). It models the unknown objective function \(f(x)\) by providing a posterior distribution given the observed data. With its simplest form, the resulting prior of GP is written as:
\begin{equation}
f(x) \sim \mathcal{GP}(\mu(x), k(x, x'))
\end{equation}
where  \( \mu(x) \) is the mean function, showing the prior belief in the objective function and \( k(x, x') \) is the covariance function (or kernel).  In this study, the surrogate model was built using a Gaussian Process (GP) with a Matérn kernel, as provided by the  \texttt{scikit-optimize} library. An anisotropic kernel was used, allowing a separate length-scale to be estimated for each parameter dimension.  The optimization process was initialized using a random sampling strategy.
Once  building the surrogate model, Bayesian optimization selects the next point \( x_{\text{next}} \) to evaluate by maximizing an \textit{acquisition function}. A common acquisition function used in  \texttt{scikit-optimize}  is \textit{Expected Improvement} (EI) and is defined  as:
\begin{equation}
\text{EI}(x) = \mathbb{E} \left[ \max(f^* - f(x), 0) \right]
\end{equation}
where \( f^* \) is the best observed current value of \( f(x) \).  The function \( f(x) \) at any \( x \)  point is normally distributed with mean \( \mu(x) \) and variance \( \sigma^2(x) \), and the improvement is modeled as:
\begin{equation}
\text{EI}(x) = (\mu(x) - f^*) \Phi\left(\frac{\mu(x) - f^*}{\sigma(x)}\right) + \sigma(x) \phi\left(\frac{\mu(x) - f^*}{\sigma(x)}\right)
\end{equation}
where \( \Phi(\cdot) \) is the cumulative distribution function (CDF) and  \( \phi(\cdot) \) is the probability density function (PDF) of the standard normal distribution, respectively. This acquisition function seeks regions where the model predicts the largest improvement over the current best value. To find the next sampling point \( x_{\text{next}} \), the acquisition function \( \text{EI}(x) \) is maximized. Once a new sample \( x_{\text{next}} \) is selected and the objective function \( f(x_{\text{next}}) \) is evaluated, the surrogate model (the Gaussian Process) is updated. The process continues until a stopping criterion is met. Further details of BO are documented in literature \cite{frazier2018tutorialbayesianoptimization}. In preliminary tests, threshold-based stopping led to extended runtimes without significant gains in accuracy. Therefore, we adopted a fixed-iteration approach of 100 iterations for practical and computational efficiency. The objective function to be minimized is defined as the summation of the root mean square error (RMSE) between experimental (\textit{in-vitro}) and numerical (\textit{in-silico}) total reaction force values for the three loading scenarios. It is given by:
\begin{equation}\label{eq:f(x)}
\text{\(f(x)\)} = \sum_{j=1}^{3} \sqrt{\frac{1}{n_j} \sum_{i=1}^{n_j} \left[R_{ij} - 2\mathcal{S}{(u_{ij}, x})\right]^2 }
\end{equation}
where \( R_{ij} \) and \(\mathcal{S}{(u_{ij}, x})\) are the experimental and numerical reaction forces, respectively, for the parameter set $x$  at displacement \(u_{ij}\), and data point \textit{i}. The term $n_j$ denotes the number of  data points for test loading \textit{j}, where \textit{j} = 1 corresponds to the one-step relaxation,  \textit{j} = 2 to the two-step relaxation, and \textit{j} = 3 to the incremental cyclic loading scenario. The factor of 2 in the equation arises from the symmetric boundary conditions modeled in the finite element (FE) simulation. Simply, the \((R_{ij}, u_{ij})\) are \textit{in-vitro} observation pairs, and  \( \mathcal{S}(u_{ij}, x) \) is the unknown force function that extracts the corresponding data from the FE simulator. The BO tries to find the model parameters $x$ that minimize the function \( f(x) \)  by repeatedly evaluating the function \( \text{EI}(x) \), which  calls the simulator (FEBio) to solve the black-box  problem. The parameter search ranges used during BO are given in Table ~\ref{table_1}, and the initial search ranges for these parameters are predicted as follows. First, the viscoelastic characteristics $g_i$ and $\tau_i$ are estimated by fitting on some trial one-step relaxation measurements using the Levenberg-Marquardt method. Then, the bounds of the estimated parameters are extended and used as search ranges for BO. Similar search ranges are considered for the fiber and matrix components. For the approximate prediction of the modulus, we initially evaluated it by analyzing the linear region of a typical incremental cyclic test loading measurement and found that the modulus of elasticity for the composite tendon (matrix + fiber) is around 350 MP. Although, to the best of our knowledge, no study has experimentally quantified the elastic modulus of the fiber and elastic component in the human Achilles tendon separately, it has been stated in some rat studies that the fiber-to-matrix ratio is between 5 and 10 \cite{dutov_measurement_2016,khayyeri_fibre-reinforced_2015, lewis_modeling_1997}. 

Sensitivity and feature importance analyzes are performed to investigate the influence of each parameter on the FRVE model for each test loading scenario. Feature importance is determined using the random forest algorithm implemented in \texttt{scikit-learn}, which is based on impurity reduction. The importance of a feature is computed as the total reduction of the criterion observed by the feature, known as the Gini importance \cite{breiman_classificatin}.  The importance of a feature $I(x_i)$ is given by: \begin{equation} I(x_i) = \frac{1}{n_t}\sum_{1}^{n_t}\sum_{s\in S_t(x_i)}\Delta I_s \end{equation} where $n_t$ is the number of trees in the forest, $S_t(x_i)$ is the set of all splits for the feature $x_i$ in the tree $t$, and $\Delta I_s$ is the decrease in the splitting criteria at split $s$. The final importance values are then normalized to 100 over all features. More details on the random forest algorithm are documented in the literature \cite{breiman_classificatin}.

As an addendum to the feature importance analysis, we implemented Sobol sensitivity analysis to investigate the sensitivity of the input parameters on the model output using \texttt{salib} \cite{herman_salib_2017}. It is a variance-based approach to quantify the parameter's sensitivity as well as their contribution to the model. Basically, it decomposes the total variance of the output \(Y\) into contributions that are attributable to input parameters and their interactions. 
Let us assume that \( Y = f(X_1, X_2, \dots, X_d) \) represents the model output as a function of \( d \) input parameters \( X_1, X_2, \dots, X_d \). The total variance of \( Y \) is given by:
\begin{equation}
V = \mathrm{Var}(Y) = \int \left( f(\mathbf{X}) - E[Y] \right)^2 p(\mathbf{X}) \, d\mathbf{X}
\end{equation}
where \( p(\mathbf{X}) \) shows the joint probability density function for the inputs. The variance can be decomposed as:
\begin{equation}
V = \sum_{i=1}^d V_i + \sum_{1 \leq i < j \leq d} V_{ij} + \cdots + V_{12\ldots d}
\end{equation}
where \( V_i = \mathrm{Var} \left( E[Y \mid X_i] \right) \) represents the main effect of input \( X_i \), and \( V_{ij} = \mathrm{Var} \left( E[Y \mid X_i, X_j] \right) - V_i - V_j \) shows the interaction effect between \( X_i \) and \( X_j \). Higher-order terms denote interactions between more input parameters. 
The Sobol indices are dimensionless quantities that measure the relative contribution of each term to the total variance. The first-order sensitivity index is given by: 
\begin{equation}
S_i = \frac{V_i}{V}.
\end{equation}
The first-order index shows the proportion of variance that is attributed to \( X_i \). Total-order sensitivity index is given by: 
\begin{equation}
S_{T_i} = 1 - \frac{V_{\sim i}}{V}
\end{equation}
where \( V_{\sim i} \) is the variance of the output excluding \( X_i \). The total-order index accounts for both the main effect and all interaction effects involving \( X_i \). The interaction terms (e.g., \( S_{ij} \)) can be calculated similarly by dividing the corresponding variance term by the total variance. However, the sensitivity due to interaction is not covered in this study.  We implemented Saltelli sampling to create the parameter space of 20,000 points for each of the 16 input parameters of the FRVE model. In our Saltelli-based Sobol sensitivity analysis, we employed uniform distributions over the parameter ranges as defined in Table ~\ref{table_1}.  The model output used for Sobol analysis was the total error function defined in Eq. \ref{eq:f(x)} which aggregates discrepancies across three experimental loading scenarios. Additional details on the Sobol sensitivity analysis are reported in the previous studies \cite{SALTELLI2010259, Saltelli2008, Iooss2015}.
\subsection{Approximate Bayesian Computation}
Although the standard Bayesian framework allows for updating prior beliefs about parameters based on the likelihood of the data given those parameters, evaluating the likelihood \( P(\mathbf{y}_{\text{obs}} | \boldsymbol{\theta}) \) is often infeasible, but the model can generate synthetic data \( \mathbf{y}_{\text{sim}} \) given parameters \(\boldsymbol{{\theta}}\). In general, Bayesian inference provides a principled framework for quantifying uncertainty in model parameters by combining prior knowledge with evidence from observed data through the likelihood function. However, for many nonlinear models, deriving an analytical or tractable likelihood function is difficult or impossible. Here, we implemented likelihood-free approximate Bayesian computation (ABC) for  stochastic calibration and uncertainty quantification of posterior parameters using  \texttt{PyABC} \cite{schalte_pyabc_2022}. The framework is based on an executable ``black-box" forward process model that simulates data using given model parameters \cite{Price02012018, wilkonson_2013}. In ABC, the computation of the likelihood is replaced by the comparison between observed (\textit{in-vitro} measurements) \(\mathbf{y}_{\text{obs}}\) and simulated data (\textit{in-silico} simulations) \( \mathbf{y}_{\text{sim}} \). Given the prior distribution of \(P(\boldsymbol{\theta})\) of  parameters \(\boldsymbol{\theta}\), the target is to approximate the posterior distribution \( P(\boldsymbol{\theta} | \mathbf{y}_{\text{obs}}) \) given observed data \( \mathbf{y}_{\text{obs}} \), a model \( f(\mathbf{y}_{\text{sim}} | \boldsymbol{\theta}) \), and a prior distribution \( P(\boldsymbol{\theta}) \).  Typically, it is based on generating simulated data \( \mathbf{y}_{\text{sim}} \) for parameters \( \boldsymbol{\theta} \) drawn from the prior distribution and retaining parameters for which \( \mathbf{y}_{\text{sim}} \) is sufficiently close to \( \mathbf{y}_{\text{obs}} \). The posterior is approximated as:
\begin{equation}
\label{epsthreshold}
P(\boldsymbol{\theta} | \mathbf{y}_{\text{obs}}) \approx P(\boldsymbol{\theta}) \cdot \mathbb{I}(d(\mathbf{y}_{\text{obs}}, \mathbf{y}_{\text{sim}}) \leq \epsilon)
\end{equation}
where \( d(\mathbf{y}_{\text{obs}}, \mathbf{y}_{\text{sim}}) \) is a distance metric, \( \epsilon \) is a tolerance threshold, and \( \mathbb{I}(\cdot) \) is an indicator function that equals 1 when the condition is satisfied \(d(\mathbf{y}_{\text{obs}}, \mathbf{y}_{\text{sim}}) \leq \epsilon \) and 0 otherwise \(d(\mathbf{y}_{\text{obs}}, \mathbf{y}_{\text{sim}}) \ge \epsilon \). The output of the algorithm is a sample set of parameters from the distribution \(P(\theta|d(\mathbf{y}_{\text{obs}}, \mathbf{y}_{\text{sim}}) \leq \epsilon)\).  Direct implementation of ABC can be inefficient, particularly for small $\varepsilon$, as it reduces the number of accepted parameters. To overcome this, pyABC performs a Sequential Monte Carlo (SMC) technique that progressively refines the posterior estimation.  It facilitates robust and efficient inference for a broad spectrum of applications. Further details of the ABC model are documented in previous studies \cite{toni_simulation-based_2010}. 

In this study, the prior distributions for the model parameters are defined based on the results of specimen-specific calibrations using Bayesian optimization (Table~\ref{table_2}). Specifically, the mean and standard deviation of each parameter, computed across nine specimens for each of the three sites, are used to construct independent Gaussian priors. To account for a potential underestimation of biological variability due to the point-estimate nature of Bayesian optimization, and to ensure coverage of physiologically plausible ranges, the standard deviations are conservatively inflated by a factor of two. The distance between observed and simulated data is computed using the Euclidean distance (L2) metric. To refine parameter estimates, the tolerance \( \epsilon \) is adaptively adjusted in the ABC SMC process. The initial tolerance threshold of \( \epsilon \) is set as 0.3. We note that we selected these options by running some trials and focusing on where the simulated data \(\mathbf{y}_{\text{sim}}\) most closely matched the observed data \(\mathbf{y}_{\text{obs}}\).

\subsection{Implementation procedures}
\begin{figure*}[!ht]
\centering
\includegraphics [width=6in] {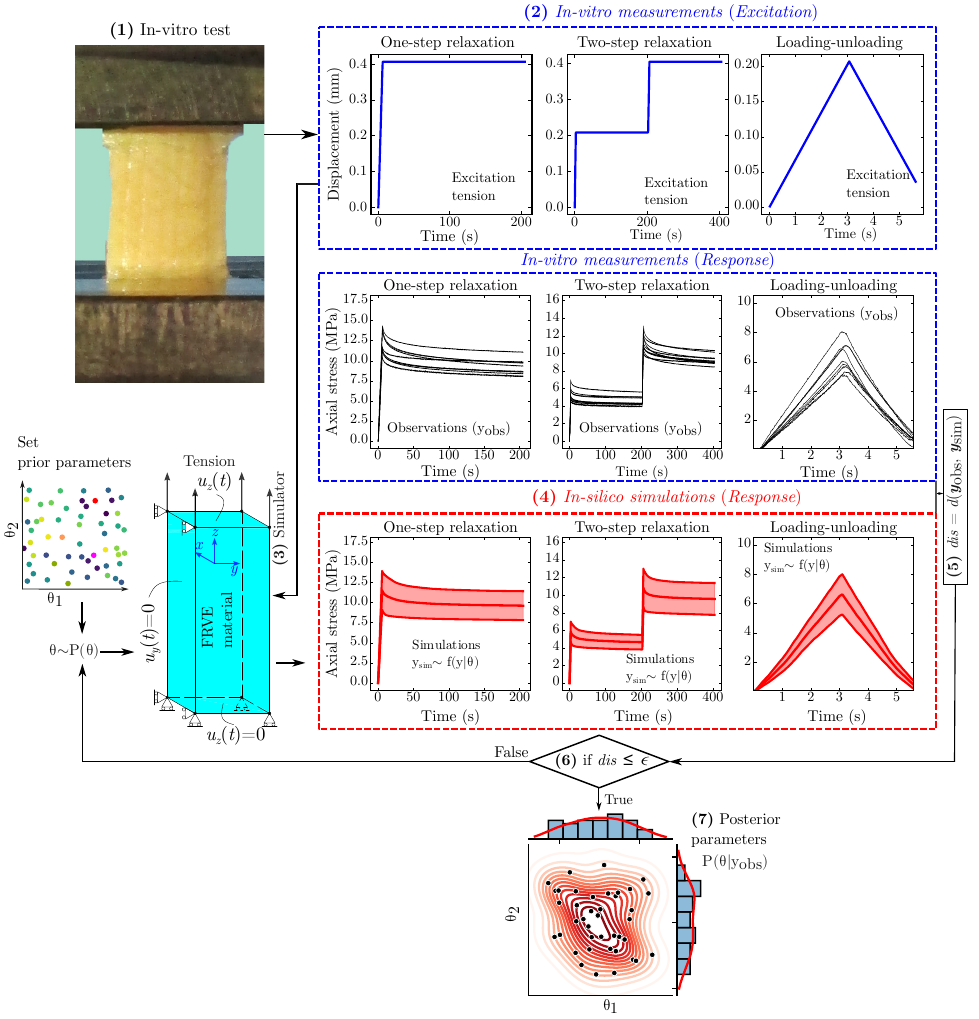}
\caption {Stochastic parameter calibration steps using \textit{in-vitro} measurements, \textit{in-silico} simulations and approximate Bayesian computation}
\label{Figure_3}
\end{figure*}

The implementation begins with the calibration of the specimen-specific parameters to determine the approximate values of the set of FRVE parameters using an optimization technique. We employed Bayesian Optimization (BO), and the procedures and stages of its implementation are illustrated in Figure ~\ref{Figure_2}, where the numbers indicate data processing steps.  Initially, test samples are harvested and re-sectioned (1). Then \textit{in-vitro} test loading protocols are applied and their measurements are acquired, categorized and saved as a serialized dataset file (2). As a preliminary study, we chose a simple uniaxial tensile test for the three test loading scenarios. For each test, the reaction force \(R(t)\) is obtained as a function of displacement \(u(t)\). This process is applied to all samples. Next, the search space is defined, and the objective function is formulated (3). \textit{In-silico} simulations are performed using the finite element method to obtain time vs. reaction force values (4). The objective function is computed as the difference between \textit{in-silico} and \textit{in-vitro} reaction force observations (5).  Once the objective function is minimized in the current step, the next target is to optimize the acquisition function, balancing exploration and exploitation (6-7). The model is then updated by incorporating a new observation and retraining the surrogate model. Finally, steps 3-7 are iterated until the stopping criteria are met (8). The implementation of \textit{in-vitro} measurements, coupled with \textit{in-silico} simulation and ABC, resembles the BO stages at the pre-processing steps. However, the main difference is that instead of evaluating specimen-specific observations, ABC includes all specimens' measurements, allowing for the stochastic calibration of parameters, including uncertainty. The implementation procedure and sequence of steps illustrated in Figure \ref{Figure_3} are as follows:
\begin{enumerate}
\item \textit{In-vitro} test samples are harvested and re-sectioned into an appropriate shape for the biomechanical test. A higher number of samples allows for better quantification.
\item \textit{In-vitro} biomechanical tests are performed to obtain output observations. Here, the resulting reaction forces \(R(t)\) are referred to as \(\mathbf{y}_{\text{obs}}\). As an addendum to the one-step and two-step relaxation loadings, we include only single-cycle observations (loading-unloading) from the incremental cyclic test during ABC inference.
\item The importance and sensitivity of each parameter are assessed using Random Forest and Sobol methods. The model parameters obtained from Bayesian optimization (BO) across all samples are then used to inform the priors in the ABC framework.
\item The next step is to perform forward simulations to drive ABC. We implemented FEBio as the simulator and created a template that includes the FRVE material model and boundary conditions that mimic the experiment. FEBio input files (\texttt{.feb}) are \(\texttt{xml}\) type files,  and we used the \texttt{jinja} package to create the FEBio input template file and serialize the data between ABC and the simulator. Once ABC determines the input parameter set \(\boldsymbol{\theta}\) that drawn from the distribution, it is passed directly to this simulator template, and the FEBio simulator generates simulated data \( f(\mathbf{y} | \boldsymbol{\theta}) \) using \(\boldsymbol{\theta}\) as: \(\mathbf{y}_{\text{sim}} \sim f(\mathbf{y} | \boldsymbol{\theta})\).
\item The distance between the observed and simulated data is computed as: \(d = d(\mathbf{y}_{\text{obs}}, \mathbf{y}_{\text{sim}})\),
where \( d(\cdot) \) is a Euclidean distance metric.
\item The parameter set \(\boldsymbol{\theta}\) is retained if the distance satisfies: \(d\leq \epsilon\); otherwise, return to step 4.
\item The joint plots of the posterior parameters are visualized, and the lower, mean and upper bounds, as well as the confidence interval 95\% of the inferred simulation \(\mathbf{y}_{\text{sim}}\), are plotted.
\end{enumerate}

\section{Results}
\subsection{Bayesian optimization}
We first performed specimen-specific parameter calibration using Bayesian optimization (BO), and the results observed for a typical specimen are represented in Figures \ref{Figure_4}, \ref{Figure_5}, \ref{Figure_6}, and \ref{Figure_7}.  {Figure \ref{Figure_4} shows that the minimum \( f(x) \) function converges rapidly, even with only a few initial iterations. This characteristic makes BO popular in machine learning for hyperparameter optimization, where the learning algorithm is treated as a black-box function. Consider the convergence plot given in Figure \ref{Figure_4}; although the initial minimum value \( f(x) \) is 65 N, after 100 iterations (or function calls), it decreases to less than 20 N. In addition, Figure \ref{Figure_5} presents the \textit{in-vitro} measurements obtained from experiments and the \textit{in-silico} observations generated using optimized FRVE parameter sets in BO. Here, we note that the terms `ground truth', `\textit{in-vitro} measurements', and `experimental observations' are often used interchangeably to convey a similar meaning. Furthermore, '\textit{in-silico} simulation', 'black-box simulator', and 'numerical observations' are also often used interchangeably to refer to a similar meaning.  Figure \ref{Figure_5} shows that the minimum \( f(x) \) values of the objective function, determined by incremental cyclic loading test measurements in one step, two steps, and incremental steps, are 5.693 N, 3.871 N, and 13.232 N, respectively. These results show that the FRVE model can capture the mechanical response of the tendon.  

We performed the BO process for 100 calls of the black-box simulator, and the parameter sensitivity can probably be improved once the number of iterations is increased. However, this can increase computational time, and we have not focused on characterizing the parameters with a high rank of sensitivity. The target of using BO is to determine how the specimen-specific parameters vary across specimens and to use these parameters to predict the priors for ABC inference. We present some parameters for the pairs of matrix and fiber constituents that were sampled during BO updating as scatter (Figure \ref{Figure_6}) and contour (Figure \ref{Figure_7}) plots to visualize where they are mainly clustered. For example, consider the sample points of \( E_m \) and \( E_f \); the elastic modulus of the matrix and fiber components is predominantly updated to around 60 MPa and 240 MPa, respectively. The remaining parameters for a typical specimen-specific calibration are given as additional figures (See Supplementary \texttt{Figures S1 and S2}). We then automated the process to characterize the remaining 26 samples, and the overall results are given in Table \ref{table_2}. These results suggest that inter-subject or inter-specimen variability and uncertainty in the biomechanical properties of these materials are high.

 \begin{figure*}[!ht]
\centering
\includegraphics [width=4in] {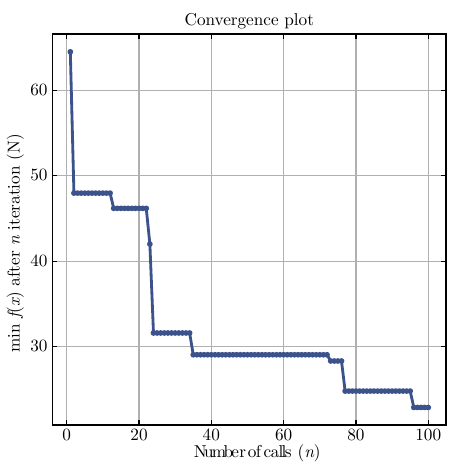}
\caption {Typical convergence plot of a specimen during Bayesian optimization}
\label{Figure_4}
\end{figure*}

\begin{figure*}[!ht]
\centering
\includegraphics [width=6in]{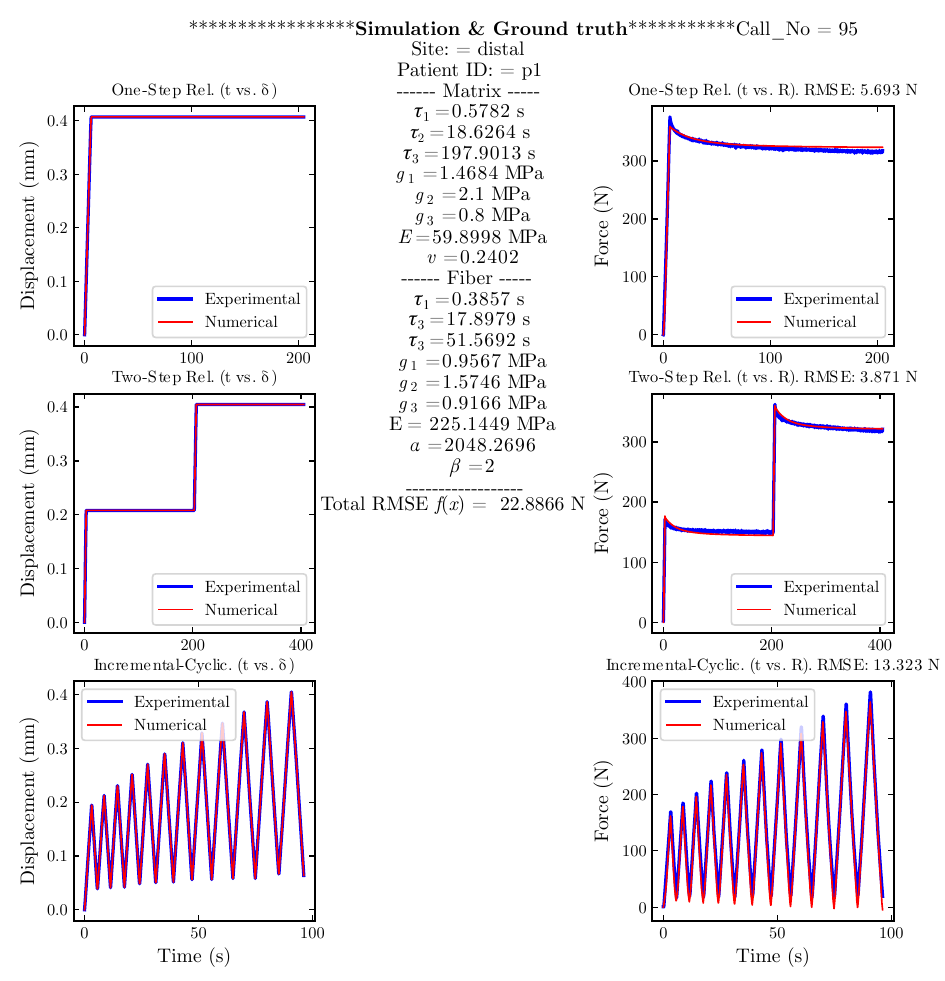}
\caption {\textit{In-vitro} (ground truth) and \textit{in-silico} (simulation) observations of a typical sample with parameters tuned with Bayesian optimization.}
\label{Figure_5}
\end{figure*}

\begin{figure*}[!ht]
\centering
\includegraphics [width=6in] {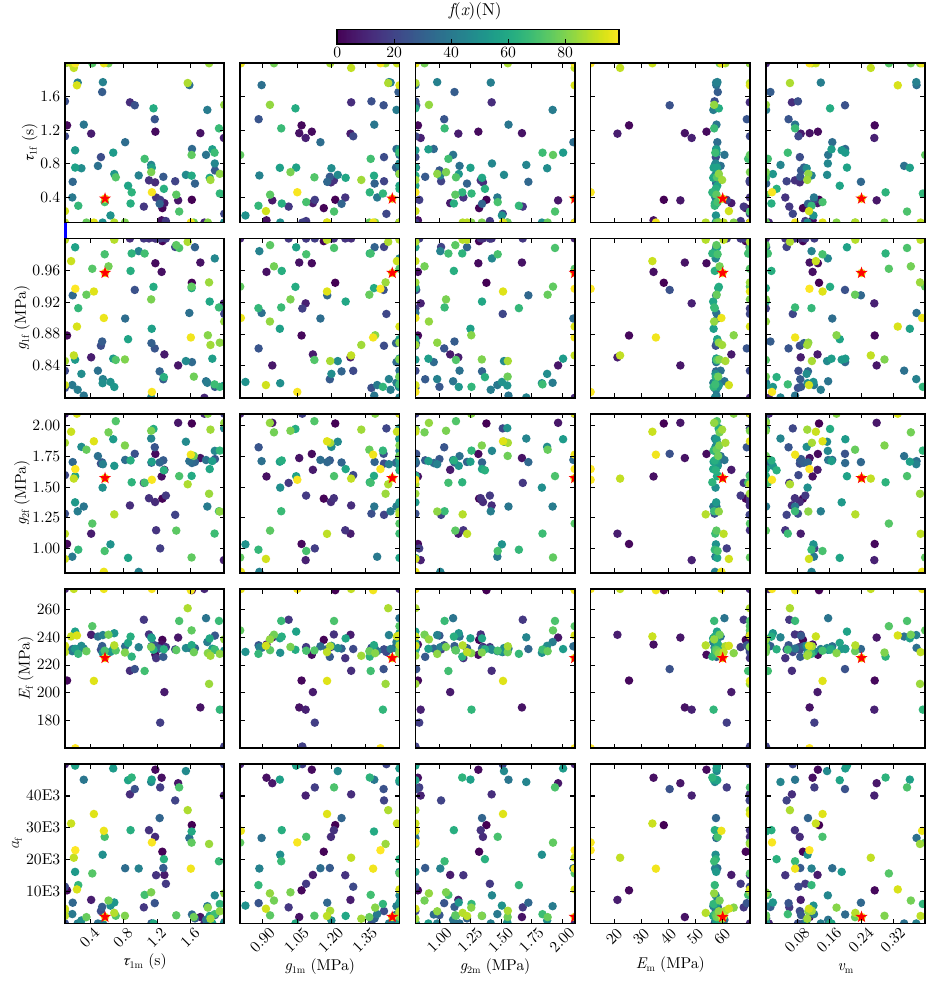}
\caption {Visualization of the sample points for some parameters of a typical specimen during Bayesian optimization. The red star indicates the best-found parameters.)}
\label{Figure_6}
\end{figure*}

\begin{figure*}[!ht]
\centering
\includegraphics [width=6in] {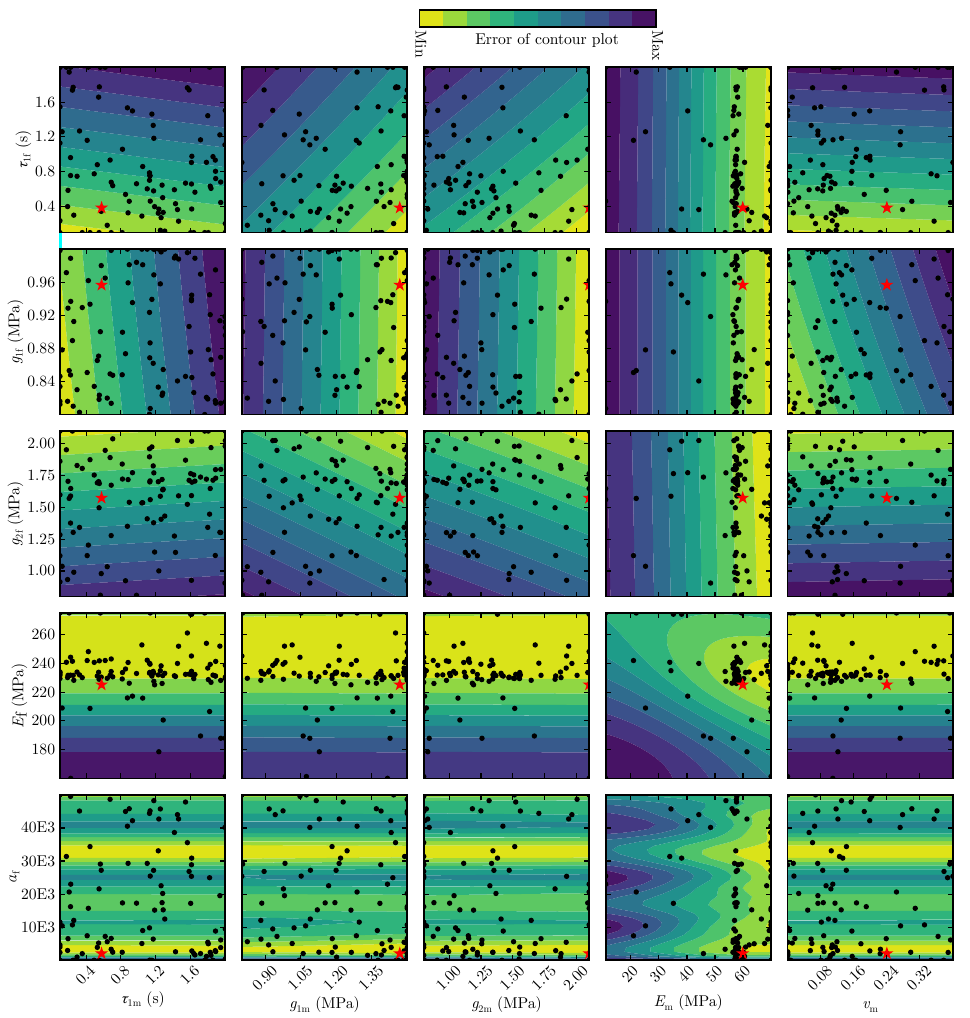}
\caption {The influence of each search space on the objective function for some parameters observed in a typical specimen during Bayesian optimization. The red star indicates the best-found parameters}
\label{Figure_7}
\end{figure*}

\begin{table}[ht]
\caption{FRVE model parameters tuned with Bayesian optimization \label{table_2}}
\centering
\small
\fontsize{8}{11}
\selectfont
\rotatebox{90}{
\begin{tabular}
{p{1.1cm} p{0.2cm} p{0.8cm} p{0.8cm} p{0.8cm} p{0.8cm} p{0.8cm} p{0.8cm} p{0.8cm} p{0.8cm} p{0.8cm} p{0.8cm} p{0.8cm} p{0.8cm} p{0.8cm} p{0.8cm} p{0.8cm} p{1cm} p{1.1cm}}
\hline
Site & ID & $\tau_{\text{1m}}$

(s) &  $\tau_{{2\text{m}}}$

(s) &  $\tau_{{3\text{m}}}$

(s) & \textit{g}\textsubscript{1m} (MPa) & \textit{g}\textsubscript{2m}

(MPa) & \textit{g}\textsubscript{3m} (MPa) & \textit{E}\textsubscript{m}

(MPa) & \textit{v}\textsubscript{m} & $\tau_{\text{1f}}$

(s) & $\tau_{\text{2f}}$

(s) & $\tau_{\text{3f}}$

(s) & \textit{g}\textsubscript{1f} (MPa) & \textit{g}\textsubscript{2f}

(MPa) & \textit{g}\textsubscript{3f}

(MPa) & \textit{E}\textsubscript{f} (MPa) & $\alpha$\textsubscript{f} & RMSE (N) \\

\hline
\multirow{9}{*}{Distal} & 1 & 0.578 & 18.626 & 197.901 & 1.468 & 2.100 & 0.800 & 59.900 & 0.240 & 0.386 & 17.898 & 51.569 & 0.957 & 1.575 & 0.917 & 225.145 & 2.05E3 & 22.887 \\
 & 2 & 0.407 & 4.937 & 60.607 & 1.315 & 1.760 & 1.944 & 70.000 & 0.254 & 1.644 & 15.729 & 83.984 & 0.943 & 1.629 & 2.072 & 206.852 & 1.09E3 & 21.352 \\
 & 3 & 0.156 & 14.878 & 152.518 & 1.463 & 0.812 & 0.981 & 62.716 & 0.300 & 1.549 & 8.717 & 20.000 & 0.905 & 1.155 & 0.800 & 216.049 & 1.30E3 & 20.892 \\
 & 4 & 1.138 & 17.810 & 146.273 & 1.461 & 1.715 & 2.100 & 31.571 & 0.240 & 2.000 & 20.000 & 133.023 & 1.000 & 2.100 & 2.100 & 203.115 & 0.00E0 & 26.799 \\
 & 5 & 0.819 & 9.548 & 194.489 & 1.352 & 1.933 & 0.800 & 58.204 & 0.121 & 2.000 & 10.558 & 40.376 & 0.905 & 1.245 & 2.100 & 244.672 & 8.62E2 & 24.957 \\
 & 6 & 0.905 & 18.524 & 23.824 & 1.500 & 1.795 & 0.800 & 34.364 & 0.001 & 2.000 & 18.802 & 169.058 & 0.821 & 2.100 & 2.100 & 167.979 & 2.23E2 & 21.311 \\
 & 7 & 2.000 & 11.041 & 142.566 & 1.500 & 2.100 & 2.100 & 70.000 & 0.400 & 2.000 & 20.000 & 200.000 & 1.000 & 2.100 & 2.100 & 229.886 & 0.00E0 & 14.737 \\
 & 8 & 2.000 & 3.000 & 113.385 & 1.472 & 1.602 & 2.100 & 70.000 & 0.228 & 2.000 & 20.000 & 20.000 & 0.913 & 2.100 & 2.100 & 182.576 & 0.00E0 & 25.667 \\
 & 9 & 2.000 & 20.000 & 20.000 & 1.500 & 2.100 & 2.100 & 46.866 & 0.000 & 2.000 & 20.000 & 200.000 & 1.000 & 2.100 & 2.100 & 167.365 & 0.00E0 & 17.196 \\
 \hline
\multirow{9}{*}{Middle} & 1 & 2.000 & 3.000 & 20.000 & 0.800 & 1.678 & 2.052 & 63.561 & 0.400 & 1.347 & 20.000 & 196.344 & 0.952 & 1.973 & 2.100 & 160.000 & 0.00E0 & 21.989 \\
 & 2 & 0.921 & 3.616 & 153.302 & 0.800 & 1.162 & 2.038 & 70.000 & 0.00 & 0.954 & 20.000 & 200.000 & 0.890 & 2.100 & 2.100 & 195.264 & 2.86E4 & 29.958 \\
 & 3 & 1.548 & 6.478 & 20.000 & 1.500 & 0.817 & 2.100 & 22.412 & 0.352 & 2.000 & 20.000 & 164.040 & 0.912 & 2.100 & 2.100 & 201.470 & 1.04E4 & 9.382 \\
 & 4 & 2.000 & 20.000 & 200.000 & 1.500 & 2.100 & 2.100 & 70.000 & 0.400 & 2.000 & 20.000 & 199.514 & 1.000 & 2.100 & 2.100 & 164.599 & 0.00E0 & 41.498 \\
 & 5 & 1.970 & 13.086 & 69.866 & 1.168 & 1.259 & 2.049 & 11.000 & 0.091 & 1.689 & 20.000 & 200.000 & 0.800 & 2.100 & 2.100 & 199.684 & 0.00E0 & 23.908 \\
 & 6 & 2.000 & 3.000 & 200.000 & 1.500 & 2.100 & 2.100 & 36.661 & 0.400 & 2.000 & 20.000 & 200.000 & 0.986 & 2.100 & 2.100 & 183.405 & 0.00E0 & 27.924 \\
 & 7 & 2.000 & 20.000 & 200.000 & 1.405 & 2.100 & 2.100 & 45.681 & 0.400 & 2.000 & 20.000 & 131.544 & 1.000 & 2.100 & 2.100 & 160.000 & 0.00E0 & 36.654 \\
 & 8 & 2.000 & 15.718 & 200.000 & 1.500 & 2.100 & 2.100 & 27.279 & 0.248 & 1.920 & 20.000 & 200.000 & 1.000 & 2.100 & 2.100 & 183.405 & 0.00E0 & 15.228 \\
 & 9 & 1.116 & 16.968 & 125.238 & 1.500 & 0.800 & 2.096 & 11.000 & 0.026 & 0.789 & 6.431 & 113.803 & 0.800 & 2.100 & 2.100 & 188.961 & 2.61E2 & 19.570 \\
 \hline
\multirow{9}{*}{Proximal} & 1 & 0.651 & 6.536 & 45.061 & 1.473 & 0.815 & 2.100 & 25.626 & 0.100 & 1.811 & 20.000 & 79.971 & 0.952 & 1.819 & 2.100 & 271.839 & 1.47E3 & 17.278 \\
 & 2 & 1.693 & 12.671 & 200.000 & 0.985 & 1.974 & 2.100 & 70.000 & 0.001 & 2.000 & 20.000 & 200.000 & 1.000 & 1.728 & 2.100 & 174.056 & 0.00E0 & 27.972 \\
 & 3 & 2.000 & 20.000 & 200.000 & 1.084 & 2.100 & 2.100 & 70.000 & 0.094 & 2.000 & 20.000 & 193.850 & 0.966 & 2.100 & 2.100 & 188.282 & 0.00E0 & 41.157 \\
 & 4 & 2.000 & 20.000 & 200.000 & 1.500 & 2.100 & 2.100 & 70.000 & 0.400 & 2.000 & 20.000 & 200.000 & 0.997 & 2.100 & 2.100 & 236.762 & 0.00E0 & 35.313 \\
 & 5 & 1.554 & 20.000 & 108.220 & 1.497 & 2.100 & 1.675 & 18.517 & 0.323 & 2.000 & 20.000 & 145.985 & 0.953 & 2.100 & 2.100 & 230.546 & 0.00E0 & 29.566 \\
 & 6 & 1.001 & 4.105 & 188.331 & 1.500 & 2.100 & 1.570 & 70.000 & 0.305 & 1.920 & 20.000 & 55.307 & 0.800 & 1.507 & 2.100 & 227.843 & 4.33E2 & 20.384 \\
 & 7 & 2.000 & 20.000 & 200.000 & 1.500 & 2.100 & 2.100 & 70.000 & 0.001 & 2.000 & 20.000 & 200.000 & 1.000 & 2.100 & 2.100 & 275.000 & 0.00E0 & 43.663 \\
 & 8 & 1.722 & 5.033 & 147.013 & 1.298 & 1.288 & 2.043 & 49.906 & 0.237 & 1.693 & 14.746 & 85.025 & 0.974 & 1.692 & 1.738 & 191.018 & 1.91E3 & 18.428 \\
 & 9 & 1.620 & 16.161 & 97.951 & 0.934 & 0.849 & 1.941 & 11.823 & 0.001 & 1.071 & 18.328 & 161.805 & 0.877 & 2.100 & 2.100 & 183.610 & 0.00E0 & 16.286 \\
 \hline

\end{tabular}
}
\end{table}

For the clinical aspect, correlation analysis between micro-scale matrix and fiber parameters calibrated via BO can also be performed. 
Here, to account for dependency due to repeated observations for the same patient's specimen harvested from distal, middle, and proximal sites, we used the \texttt{rm\_corr()} repeated measures correlation function from the Pingouin statistical package. Repeated measures reveal that the most significant correlations are observed between $\tau_{3m}$ and $\tau_{1f}$ (\textit{r} = 0.8), $g_{3m}$ and $g_{3f}$ (\textit{r} = 0.71), $\tau_{2m}$ and $\tau_{1f}$ (\textit{r} = 0.69), $g_{2m}$ and $\tau_{1f}$ (\textit{r} = 0.65), $g_{2m}$ and $g_{3f}$ (\textit{r} = 0.64), while the only negative significant correlation exists between $E_{m}$ and $\tau_{1f}$ (\textit{r} = 0.48) (\textit{p} $<$ 0.05) (Figure \ref{Figure_8}).

\begin{figure*}[!ht]
\centering
\includegraphics [width=5in] {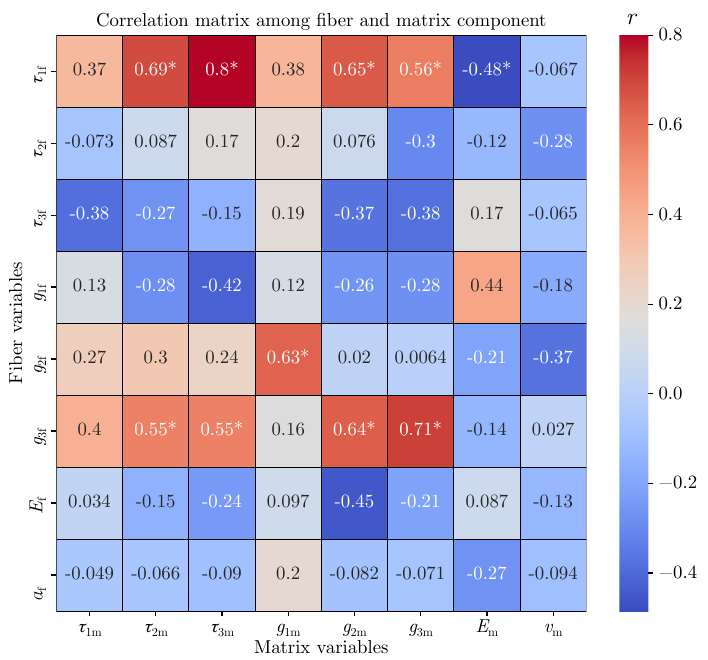}
\caption {Correlation levels between matrix and fiber variables. *Repeated measures correlation coefficient (\textit{r}) is significant at the \textit{p} $<$ 0.05 level}
\label{Figure_8}
\end{figure*}

\subsection{Sensitivity analysis}
To determine the influence of each parameter on the objective function of the FRVE model parameters, we investigate the percentage importance of the parameters using random forest and their sensitivity using Sobol. Here, we use the observations of the error function \( \min f(x) \) given in Eq. \ref{eq:f(x)} as output during sensitivity analysis and importance identification. The results of the random forest for a typical specimen in three loading conditions are represented in Figure \ref{Figure_9}. As expected, the elastic modulus of the fiber \( E_f \), the exponential coefficient \( \alpha_f \), and the elastic modulus of the matrix \( E_m \) have the highest feature importance. Interestingly, the viscoelastic characteristics \( g_{im} \) and relaxation time coefficients \( \tau_{im} \) of the ground matrix at the \( i \)th Prony order are more important than the fiber's parameters for the combined three-loading \( \min f(x) \) function.

\begin{figure*}[!ht]
\centering
\includegraphics [width=6in] {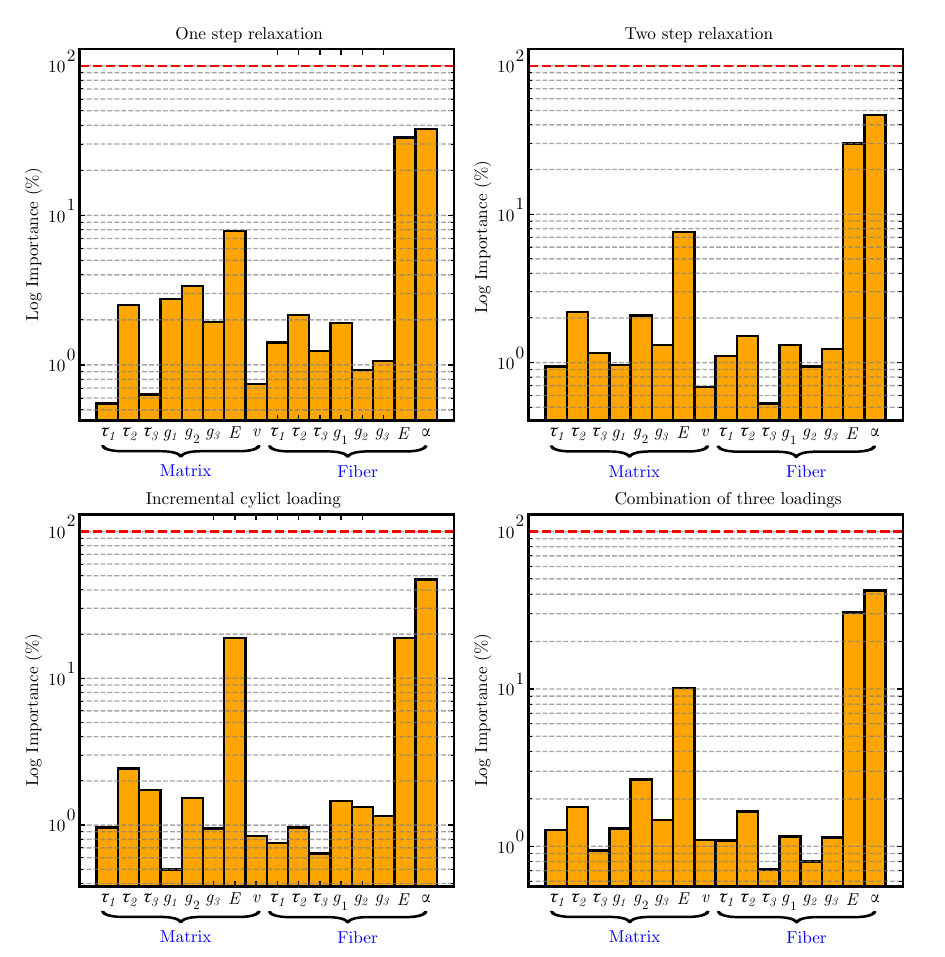}
\caption {Variables feature importance using Random forest}
\label{Figure_9}
\end{figure*}

For Sobol's sensitivity analysis, the parameter space is created using Saltelli sampling. We generate 20,000 input sampling sets, each consisting of 16 parameters (8 for the ground matrix and 8 for the fiber), and collect the output \( \min f(x) \) observations. Similar to the results obtained from the random forest analysis, \( E_f \), \( E_m \), and \( \alpha_f \) are identified as the most sensitive parameters in the model. The total and first-order sensitivity indices for these parameters are presented in Figure~\ref{Figure_10}. Based on the total-order index scores, the three most sensitive parameters, in descending order, are: \( E_f \) (0.7325), \( E_m \) (0.4905), and \( \alpha_f \) (0.3587).

\begin{figure*}[!ht]
\centering
\includegraphics [width=6in] {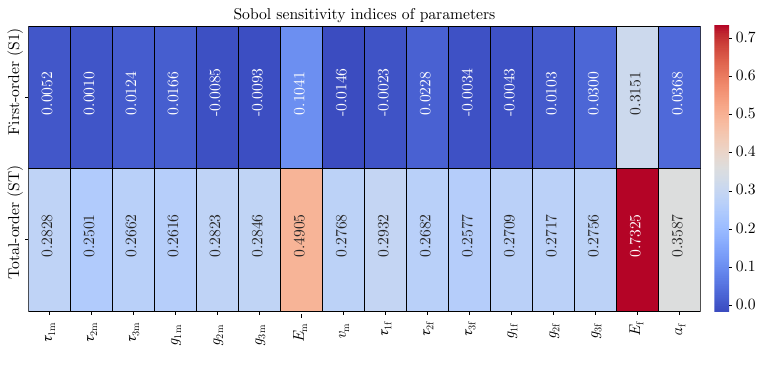}
\caption {Sensitivity of the parameters}
\label{Figure_10}
\end{figure*}

\subsection{Approximate Bayesian computation}
We presented specimen-specific parameter identification results in the previous section and observed that the variability in mechanical parameters across samples is high, potentially influenced by factors such as donor age, genetics, tissue pathology, gender, and other sample-specific characteristics. To determine the uncertainty in these parameters and stochastically calibrate them, we implement likelihood-free inference in a probabilistic framework. The ABC inference is initiated with a tolerance threshold of \( \epsilon = 0.3 \), selected empirically to allow for sufficient sampling diversity in the initial generation. This threshold is adaptively updated by the ABC-SMC algorithm in subsequent populations, progressively tightening the acceptance criterion around the observed data. In each generation, 40 accepted sample sets are retained.The inferred parameters for the distal, middle, and proximal sites are given in Table \ref{table_3}. Consider \( E_f \) at the distal site: the mean is 216.2 MPa, and the 95\% HDI bounds are 164.6 MPa and 267.8 MPa.

\begin{table}
\centering

\caption{Posterior parameters inferred for the three sites \label{table_3}}
\small
\fontsize{8}{11}

\renewcommand{\arraystretch}{1.23}  
\begin{tabular}{p{1.6cm} p{4.2cm} p{4.2cm} p{4.2cm}}
\hline
 Parameter & Distal & Middle & Proximal \\
\hline
\textit{E}\textsubscript{f} (MPa) & 216.2 ± 31.12 (164.6, 267.8) & 249.1 ± 46.47 (167.2, 324.9 & 238.5 ± 50.04 (161.4, 322.3) \\
$\alpha_\text{f}$ (-) & 515.7 ± 224.4 (105.7, 883.0)  & 566.1 ± 275.9 (76.26, 962.6) & 548.9 ± 278.9 (76.26, 962.6) \\
\textit{g}\textsubscript{1f} (MPa) & 1.305 ± 0.254 (0.813, 1.772) & 1.288 ± 0.280 (0.881, 1.762) & 1.252 ± 0.272 (0.881, 1.762) \\
\textit{g}\textsubscript{2f }(MPa) & 1.857 ± 0.602 (0.855, 2.828) & 1.726 ± 0.676 (0.803, 2.789) & 1.814 ± 0.652 (0.803, 2.789) \\
\textit{g}\textsubscript{3f} (MPa) & 1.754 ± 0.638 (1.001, 2.896) & 1.849 ± 0.629 (0.847, 2.820) & 1.741 ± 0.633 (0.847, 2.820) \\
$\tau_{\text{1f}}$ (s) & 0.988 ± 0.559 (0.116, 2.007) & 1.086 ± 0.661 (0.167, 2.077) & 1.001 ± 0.623 (0.167, 2.077) \\
$\tau_{\text{2f}}$ (s) & 11.31 ± 5.395 (2.347, 19.41) & 13.06 ± 5.395 (2.798, 21.40) & 12.79 ± 5.550 (3.023, 21.75) \\
$\tau_{\text{3f}}$ (s) & 106.0 ± 49.25 (22.91, 189.5) & 120.9 ± 57.21 (30.07, 218.5) & 119.9 ± 55.14 (24.44, 218.5) \\
\textit{E}\textsubscript{m} (MPa) & 42.16 ± 14.74 (18.26, 67.07) & 46.19 ± 18.76 (14.82, 77.67) & 42.76 ± 18.36 (14.82, 79.59) \\
\textit{g}\textsubscript{1m }(MPa) & 1.499 ± 0.348 (0.941, 2.158) & 1.626 ± 0.440 (0.929, 2.296) & 1.600 ± 0.428 (0.929, 2.296) \\
\textit{g}\textsubscript{2m} (MPa) & 1.948 ± 0.537 (0.861, 2.742) & 1.841 ± 0.600 (1.033, 2.830) & 1.752 ± 0.582 (1.012, 2.830) \\
\textit{g}\textsubscript{3m }(MPa) & 1.809 ± 0.620 (0.885, 2.834) & 1.946 ± 0.528 (1.059, 2.856) & 1.898 ± 0.553 (0.896, 2.824) \\
$\tau_{\text{1m}}$ (s) & 0.814 ± 0.447 (0.188, 1.671) & 0.977 ± 0.479 (0.264, 2.042) & 0.960 ± 0.531 (0.179, 2.042) \\
$\tau_{\text{2m}}$ (s) & 12.99 ± 5.199 (3.993, 21.96) & 12.12 ± 6.041 (3.521, 22.88) & 12.40 ± 6.176 (3.152, 22.15) \\
$\tau_{\text{3m}}$ (s) & 83.58 ± 46.82 (20.62, 192.6) & 112.1 ± 55.61 (24.08, 212.1) & 104.8 ± 53.14 (24.08, 212.1) \\
\textit{v}\textsubscript{m}   (-) & 0.208 ± 0.095 (0.069, 0.392) & 0.202 ± 0.121 (0.015, 0.399) & 0.210 ± 0.118 (0.015, 0.399) \\
\hline
\multicolumn{4}{l}{The values are represented as mean ± standard deviation.}\\
\multicolumn{4}{l}{The values given in round brackets are the highest density interval (HDI) bounds at 95\% level }\\
\end{tabular}

\end{table}

We visualized the posterior joint distributions to investigate the relationships, correlations, and uncertainty between the parameters after updating beliefs based on \textit{in vitro} measurement observations. Some parameter pairs plotted for the matrix and fiber components, as well as their distributions for the distal site, are shown in Figure \ref{Figure_11}. Consider \( g_1 \): there is a significant negative correlation between the matrix and the fiber component (\textit{r} = -0.37, \textit{p} $<$ 0.05). However, the remaining parameters shown in Figure \ref{Figure_11} for the matrix (\(\tau_{1m}\), \(g_{2m}\), \(E_{m}\), \(v_{m}\)) and the fiber (\(\tau_{1f}\), \(g_{1f}\), \(g_{2f}\), \(\alpha_{f}\)) are not significantly correlated. The joint posterior plots inferred for 16 parameters for three sites are given in Supplementary \texttt{Figures S3-S5}. Once the calibration of the parameters is complete and their uncertain ranges for the highest density interval (HDI) 95\% are inferred, the next step is to investigate how well the simulated data from the posterior \( f(y_{sim} |\theta) \) can capture the uncertainties of \textit{in vitro} observations \( y_{obs} \). We perform forward simulations based on posterior parameters (\(\theta\)) for two cases. In the first case, we performed three simulations: one based on the mean and the other two based on the low and high bounds of \(\theta\) given in Table \ref{table_3}. We then collected observations from \( y_{sim} \). The inferred observation band is shown in cyan, and its bounds are represented in blue in Figure \ref{Figure_12}. In the second case, we performed 40 simulations, each corresponding to a single set of posterior parameters, and then computed the 95\% confidence interval (CI) of the 40 \( y_{sim} \) observations. The observation band for the 95\% CI and its bounds are colored in red (Figure \ref{Figure_12}). The results obtained in both cases demonstrate that the stochastically calibrated model parameters can capture the uncertainty band of \textit{in vitro} measurements (\( y_{obs} \)). However, to characterize and simulate the soft tissue response under extreme uncertainty, it is preferable to use the 95\% HDI bound of \(\theta\), as demonstrated in case 1.

\begin{figure*}[!ht]
\centering
\includegraphics [width=6in] {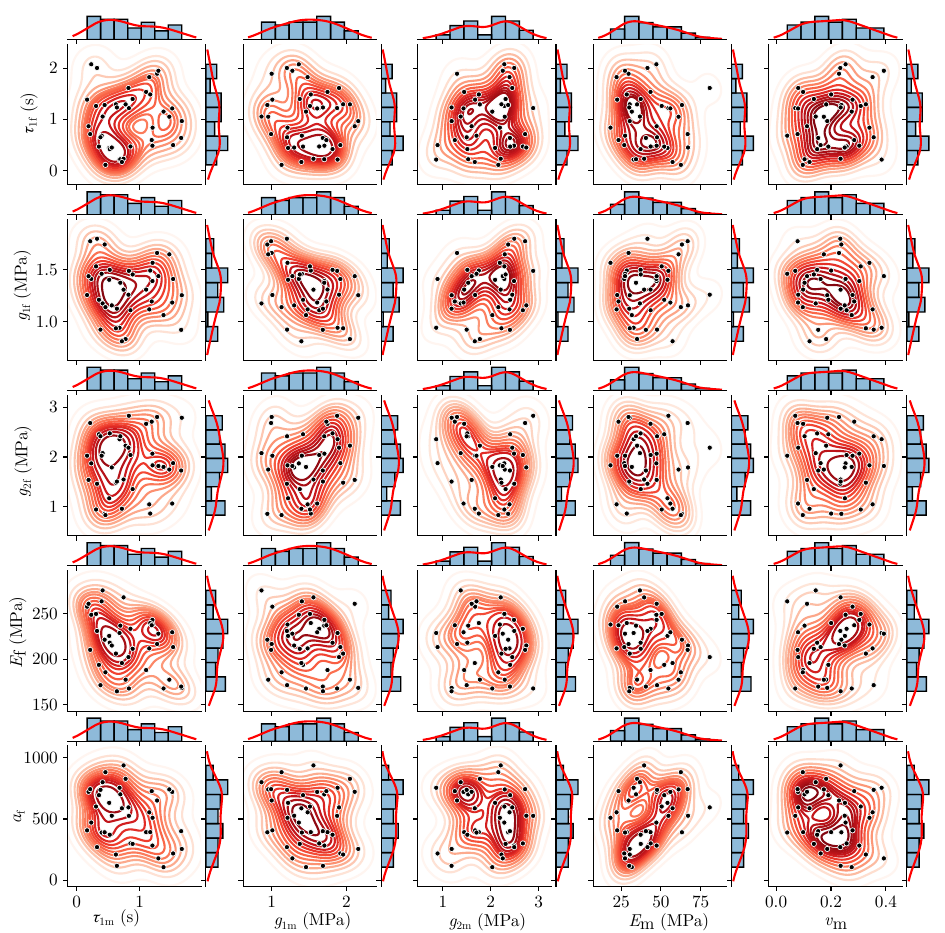}
\caption {Posterior joint plots between same parameters of the matrix and fiber component for the distal site}
\label{Figure_11}
\end{figure*}
\begin{figure*}[!ht]
\centering
\includegraphics [width=6in] {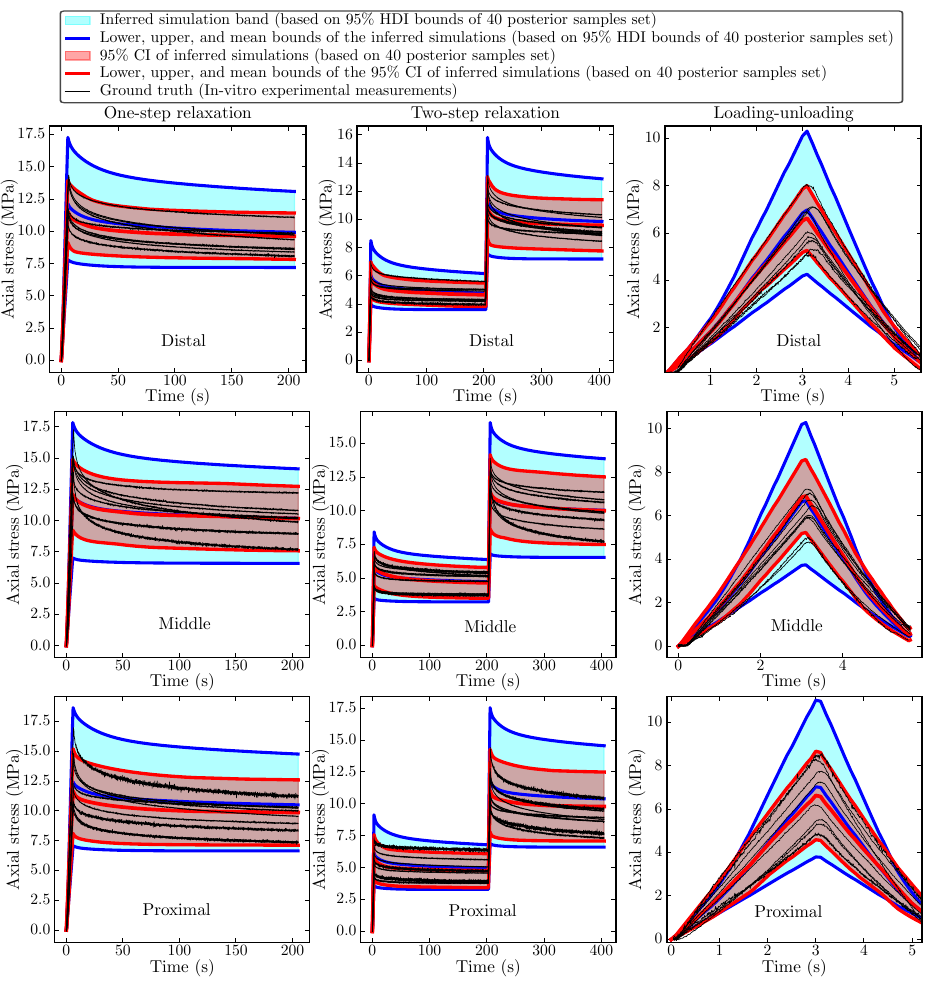}
\caption {\textit{In-vitro} measurements (ground truth) and inferred simulation observations based on posterior samples for the three sites}
\label{Figure_12}
\end{figure*}
In some clinical biomechanical studies, it may be necessary to compare the data obtained from different groups. This is usually carried out by applying parametric or non-parametric statistical tests to the results. This approach is effective within a frequentist framework. Although this study does not directly investigate the differences between the three sites, we performed unsupervised machine learning to determine whether the probabilistic data of the groups are clustered and separated. Here, we primarily used principal component analysis (PCA) and receiver operating characteristic (ROC) curves to investigate whether data from different groups could be effectively distinguished. Prior to PCA, we standardized the posteriors \( \theta \) as \( \frac{\theta - \mu}{\sigma} \), where \( \mu \) is the mean and \( \sigma \) is the standard deviation. Then, we processed 40 parameters with a 16-dimensional posterior parameter set through PCA to reduce them to a single-dimensional component using the \texttt{scikit-learn} library. The reduced component observations corresponding to a specific site were grouped in pairs and represented via ROC curves. A perfect separation between two groups has an area under the curve (AUC) of 1, while an AUC value around 0.5 indicates minimal separation. The results given in Figure ~\ref{Figure_13} suggest that the posterior parameters are not well separated between the distal vs. middle sites (AUC = 0.50), the middle vs. proximal (AUC = 0.51) and the distal vs. proximal sites (AUC = 0.51). Furthermore, the kernel density distribution exhibited comparable findings, suggesting a lack of clear separation (Figure \ref{Figure_13}). Finally, using stochastically inferred parameters from \textit{in-silico} simulations, it is also possible to investigate the stress uncertainties of the fiber and matrix constituents. A typical stress band observed for the distal site is represented in Figure \ref{Figure_14}, and for the middle and proximal results are given in Supplementary \texttt{Figure S6}.

\begin{figure*}[!ht]
\centering
\includegraphics [width=6in] {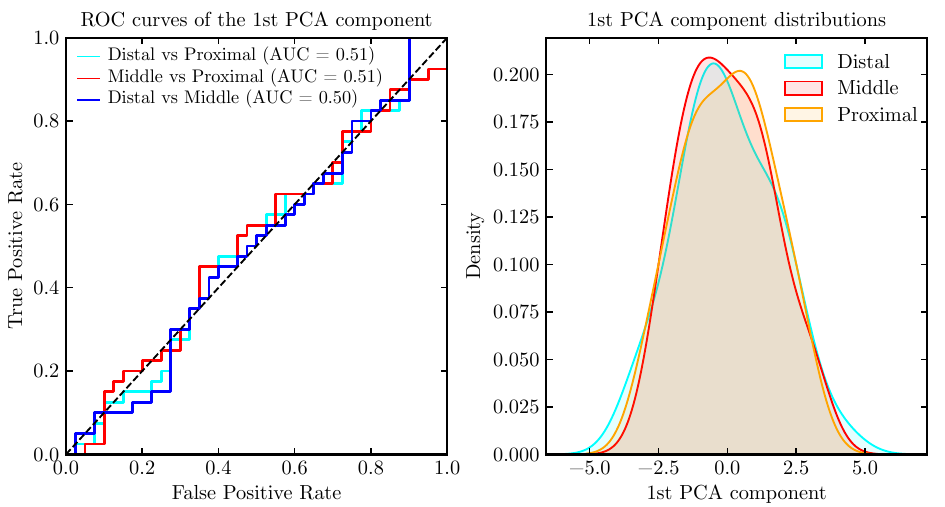}
\caption {ROC curves and kernel density estimation for the 1st PCA component}
\label{Figure_13}
\end{figure*}

\begin{figure*}[!ht]
\centering
\includegraphics [width=6in] {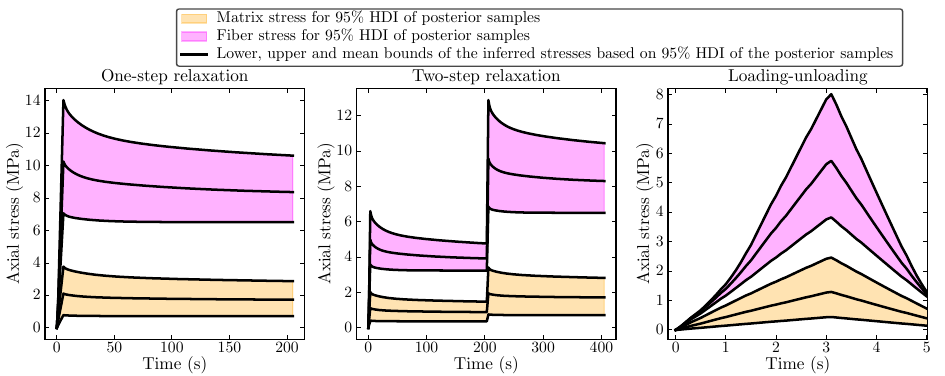}
\caption {Inferred matrix and fiber stresses for the distal site }
\label{Figure_14}
\end{figure*}

\section{Discussion and Conclusions}
\textit{In-silico} simulations using computational techniques have significant potential to model very specific cases, ranging from patient-specific tissue behavior to a large variation in population. One significant factor that influences the outcomes of the simulation is the selected constitutive material model, as well as the parameters used in this model. Homogeneous isotropic hyperelastic materials can capture the biomechanical response of many deformable solids. However, most biological materials exhibit an inherent variation in their response and may not be accurately represented by a constitutive model with a single set of deterministic parameters. In this approach, the model is typically verified through forward model evaluations using a certain number of experiments. For the backward problem, an inverse approach is mostly used to identify the input parameters from the output. This method is powerful for the characterization of patient- or specimen-specific materials. To account for the uncertainties in the experimental observations, the standard Bayesian approach is widely implemented in many studies \cite{madireddy_bayesian_2015, teferra_bayesian_2019, brewick_uncertainty_2018, staber_stochastic_2017, mihai_LA}. However, it may not always be possible to derive and express the likelihood function in explicit form, particularly for complex loading scenarios. We suggest a likelihood-free approximate Bayesian computation based on a black-box simulator. We demonstrate its implementation for Achilles tendon specimens from diabetic patients, which include deep uncertainties in their \textit{in-vitro} observations.

There are a wide variety of factors, including pathology, microstructural characteristics, tissue age, and embalming duration, that can significantly influence the biomechanical response. The goal is to stochastically calibrate parameters that represent the uncertain bounds of experimental observations without using the likelihood function. We consider the FRVE model for the representation of \textit{in-vitro} simulations with 16 parameters. Sobol sensitivity analysis and feature importance through a random forest are performed to investigate the influence of parameters on the model. The overall results we observed are as follows. First, \textit{ in vitro} measurements exhibit extreme uncertainties and variability between samples is high. Second, Bayesian optimization converges rapidly with just a few initial iterations and calibrates 16 specimen-specific parameters. Third, sensitivity analysis and feature importance showed that $E_f$, $\alpha_f$, and $E_m$ parameters have the highest influence on the model. Additionally, repeated measures correlation showed a significant correlation between the matrix and fiber viscoelastic parameters. Fourth, the results obtained from ABC demonstrated that the stochastically inferred model parameters can capture the uncertainty band of \textit{in-vitro} measurements ($y_{obs}$). Fifth, the PCA results demonstrated that the posterior parameters are not separated between sites. For clinical implementation, this observation may suggest that treatment or rehabilitation strategies regarding the mechanical response of diabetic tendons may be generalized to target the entire tendon rather than focusing on specific sites. However, additional data are needed to investigate whether these findings hold in different patient populations.

Although ABC is conceptually simple and can be easily implemented for calibration purposes, it can be computationally expensive due to the high processing time in the simulator. However, we believe that this issue can be addressed by using machine or deep learning algorithms as a surrogate model instead of obtaining \(y_{sim}\) observations from the simulator during each step of the ABC sampling update. Once the input parameter set \(\theta_N\) and the corresponding observations \(y_{sim}\) from the forward simulator are trained in the algorithm, unknown simulation observations based on the input parameters sampled in ABC can be predicted using the surrogate model.  
  
Mechanical characterization of soft tissue is usually performed by testing \textit{n} samples and then averaging the model parameters calibrated for \textit{n} data sets. In some cases, the average of \textit{n} measurements is taken first, and then non-linear curve fitting techniques for the constitutive model are applied to calibrate the parameters for a single data set. In both methods, the calibrated parameters cannot fully capture the uncertainties in the measurements. Moreover, while uncertainties arising from geometry are often not considered, the proposed method can easily incorporate these uncertainties.

This study shows that the parameters calibrated using the probabilistic technique (Table~\ref{table_3}) differ from those obtained through the frequentist approach (Table~\ref{table_2}). In the frequentist approach, we optimize parameters separately for each specimen based on its own data using Bayesian Optimization (BO). In contrast, in the ABC approach, we use data from all specimens collectively to infer posterior parameter distributions, capturing uncertainty in the estimates. For example, consider the elastic modulus of fiber \( E_f \) for the distal site: using the frequentist approach, the average \( E_f \) value for nine specimens is 204.82 ± 27.42 MPa (mean ± standard deviation), whereas the ABC estimate is 216.2 ± 31.12 MPa. While this difference illustrates how inference outcomes may vary between methods, it should be interpreted with caution due to key modeling assumptions. For instance, the initial choice of \( \epsilon = 0.3 \) was made empirically. While the ABC-SMC algorithm adaptively updates this threshold in later generations, the initial value still reflects a modeling decision that may influence the resulting posterior distributions. Furthermore, the following limitations should be addressed in future studies. First, the number of samples is important for the quantification of uncertainty; hence, more specimens should be used to investigate the posterior parameters of the population. Second, investigations are performed on samples with approximately similar geometrical dimensions. Further studies can include geometrical uncertainty. Third, while the concept of ABC is simple and widely applicable, it can be computationally expensive, as it requires repeatedly simulating the forward model (e.g., using the finite element method). Future studies could explore the use of surrogate models based on machine learning or deep learning algorithms to reduce the computational cost of the forward simulator. Finally, this study focused solely on quantifying uncertainty arising from variability in mechanical parameters across specimens with approximately similar geometry. Other important sources of uncertainty—such as measurement noise, model-form errors, and geometric variability—were not considered in the current framework. Future work could extend this approach by incorporating these additional sources of uncertainty.

In conclusion, the likelihood-free uncertainty quantification approach provides a useful framework for the stochastic calibration of constitutive material model parameters without the need to derive a likelihood function. This general framework can be applied to various combinations of loading data and constitutive models. In medicine, microscale characterization and quantification of uncertainty of constituents that cannot be easily determined using experimental techniques, including constituent parameters, constituent stress, and their correlations, could help create more effective therapeutic strategies.
\section{Acknowledgments}
The author thanks Prof. Firat Ozan from Kayseri City Hospital for his constructive comments and suggestions for this work.

\clearpage
\bibliographystyle{elsarticle-num} 
\bibliography{references}
\clearpage

\section*{Supplementary Materials}
\begin{figure*}[!ht]
\centering
\includegraphics [width=7.2in] {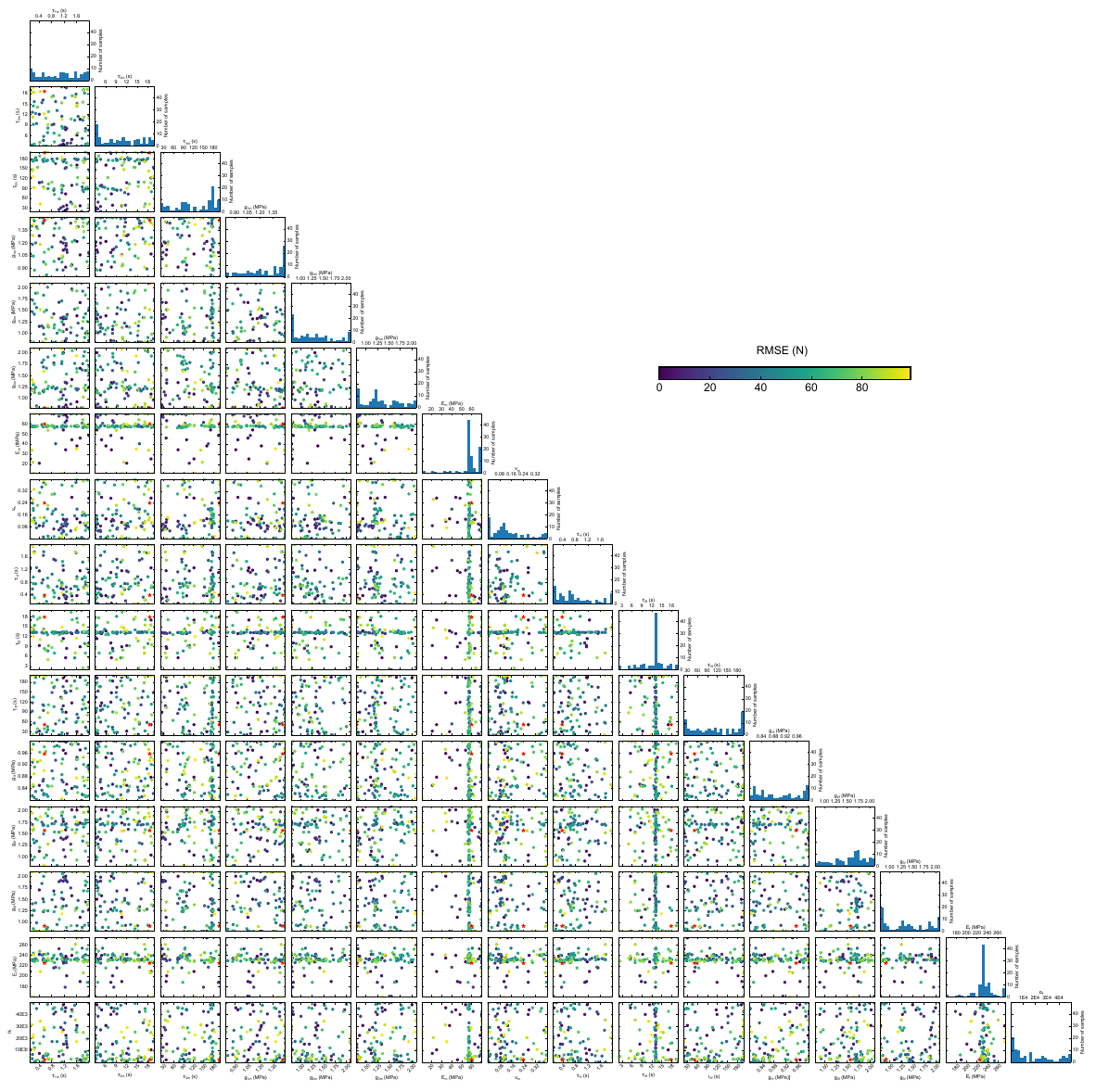}
\renewcommand{\thefigure}{S1} 
\caption {Visualization the sample points for a typical specimen during Bayesian optimization. (The red star shows the best-found parameters)}
\label{Figure_S1}
\end{figure*}

\begin{figure*}[!ht]
\centering
\includegraphics [width=7.2in] {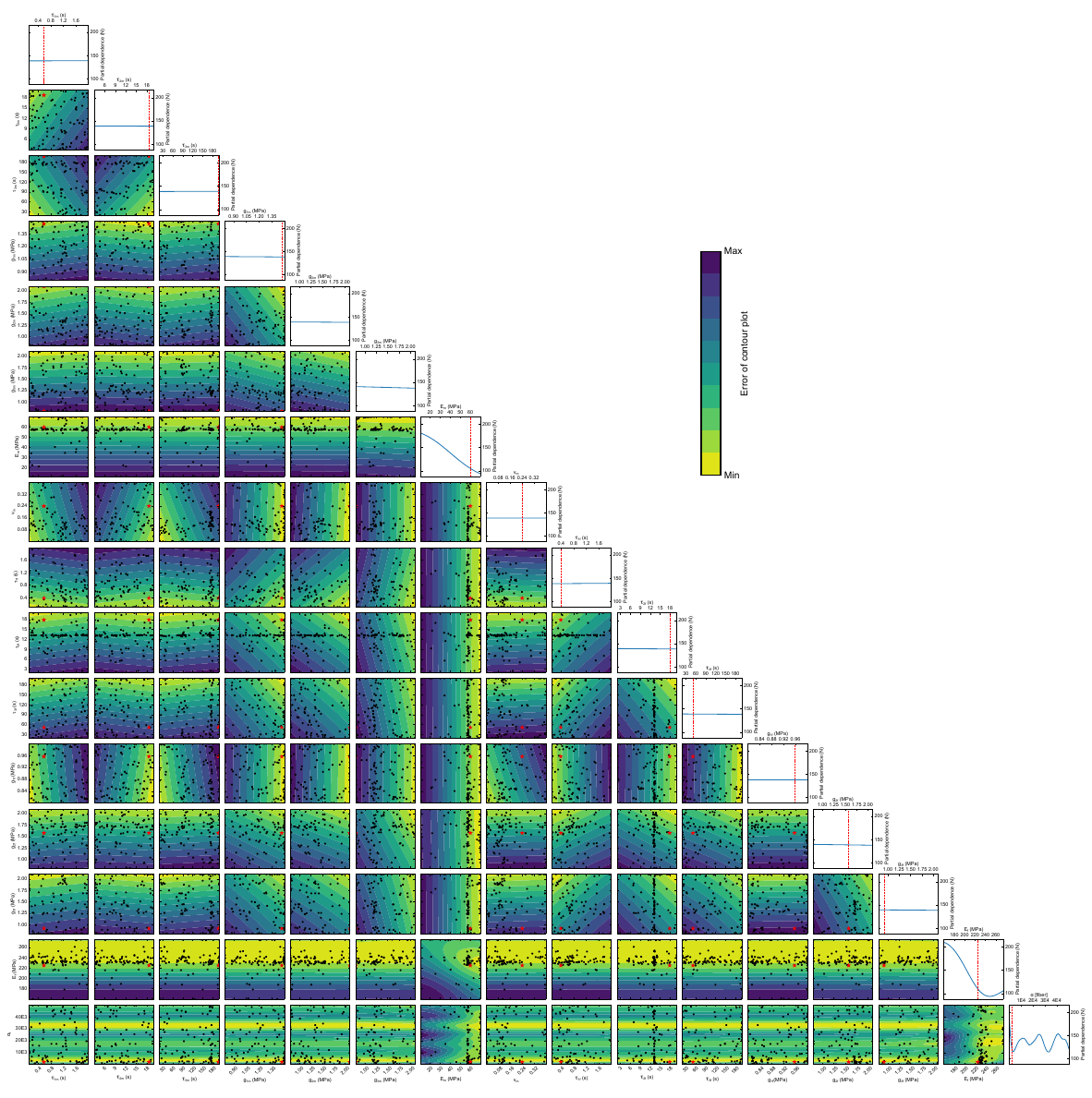}
\renewcommand{\thefigure}{S2} 
\caption {The influence of each search space on objective function observed for a typical
specimen in Bayesian optimization (The red star shows the best-found parameters}
\label{Figure_S2}
\end{figure*}

\begin{figure*}[!ht]
\centering
\includegraphics [width=7.2in] {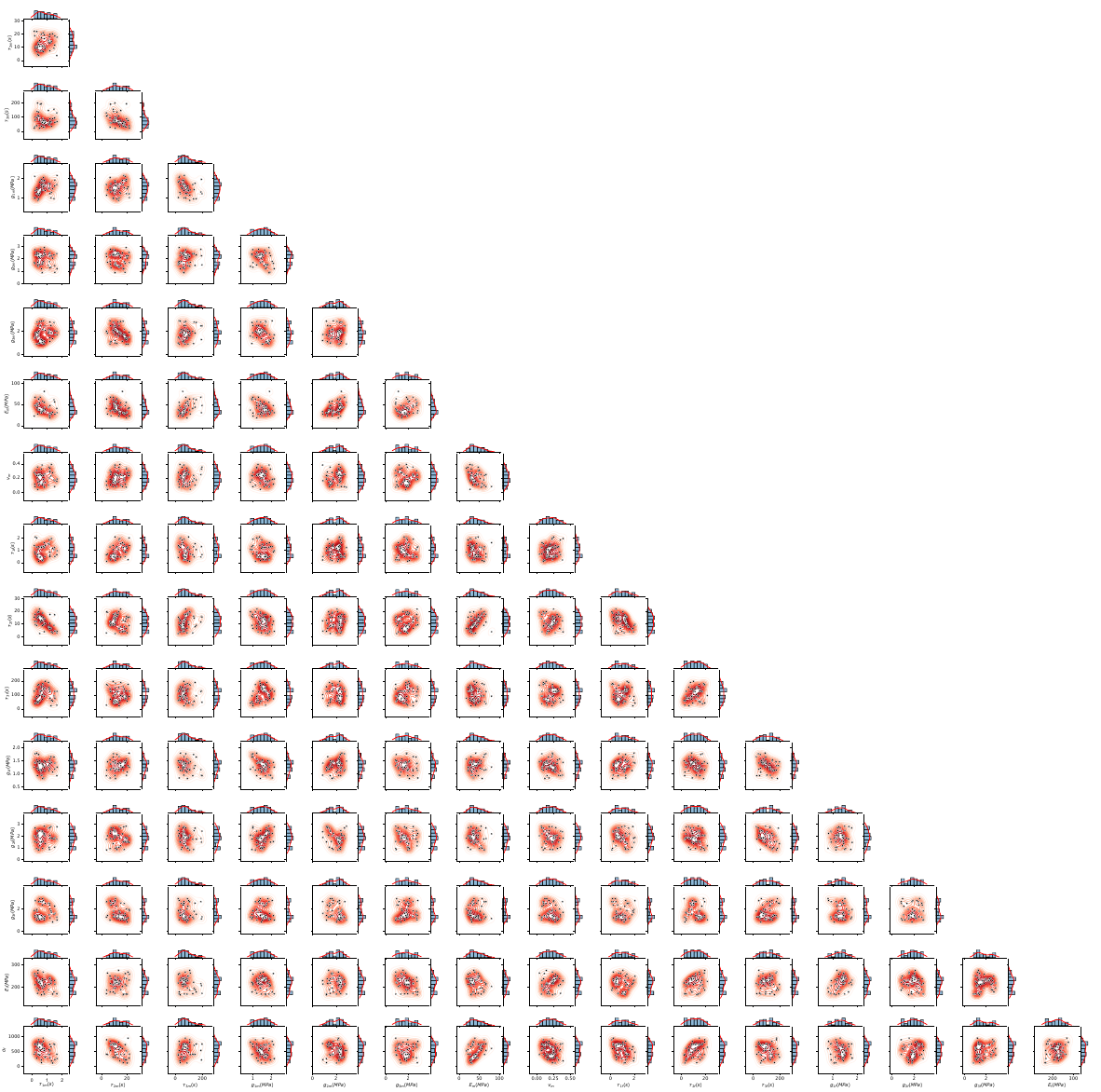}
\renewcommand{\thefigure}{S3} 
\caption {Posterior joint plots between parameters of the matrix and fiber component
for the distal site}
\label{Figure_S3}
\end{figure*}

\begin{figure*}[!ht]
\centering
\includegraphics [width=7.2in] {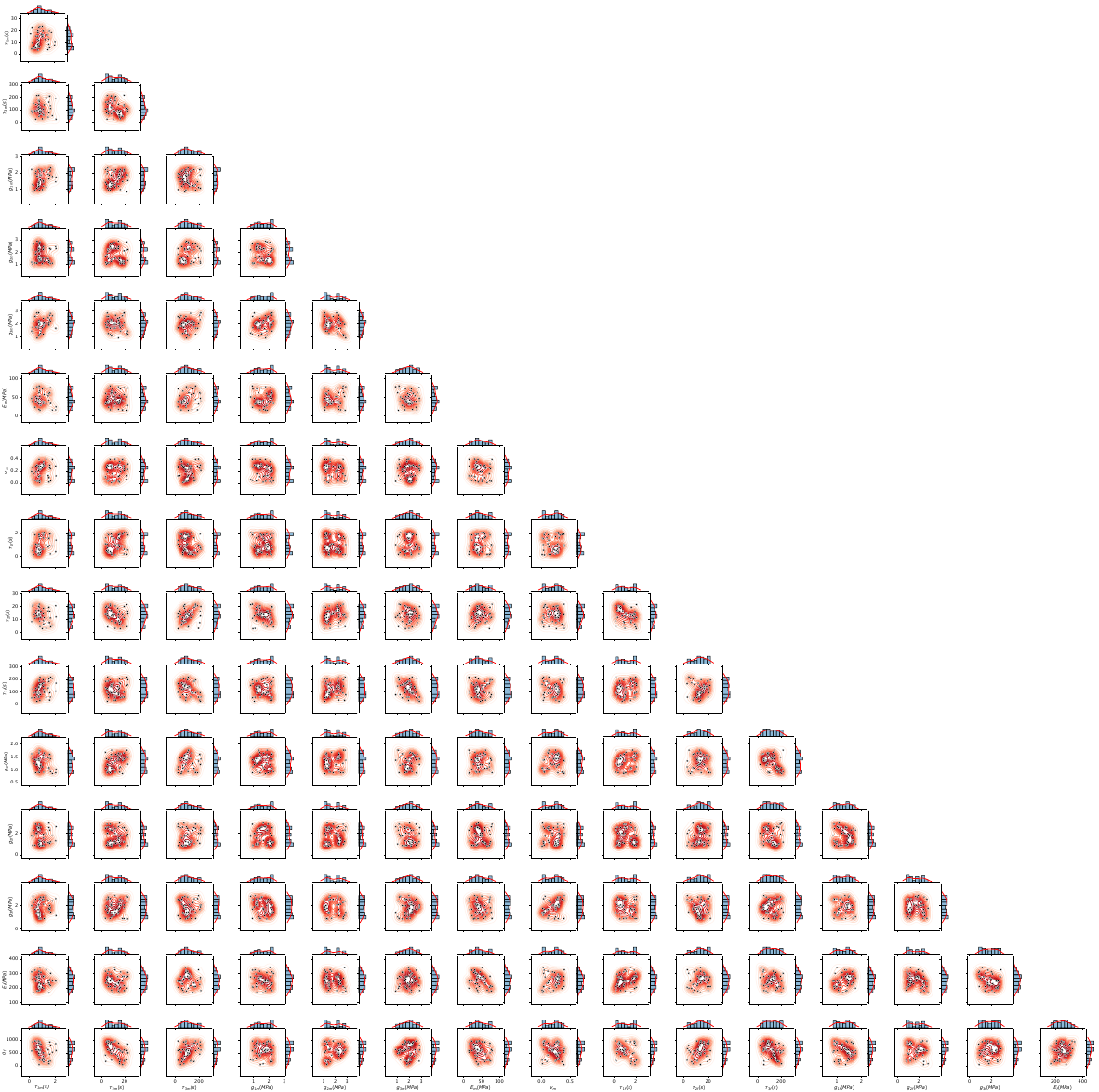}
\renewcommand{\thefigure}{S4} 
\caption {Posterior joint plots between parameters of the matrix and fiber component for the middle site}
\label{Figure_S4}
\end{figure*}

\begin{figure*}[!ht]
\centering
\includegraphics [width=7.2in] {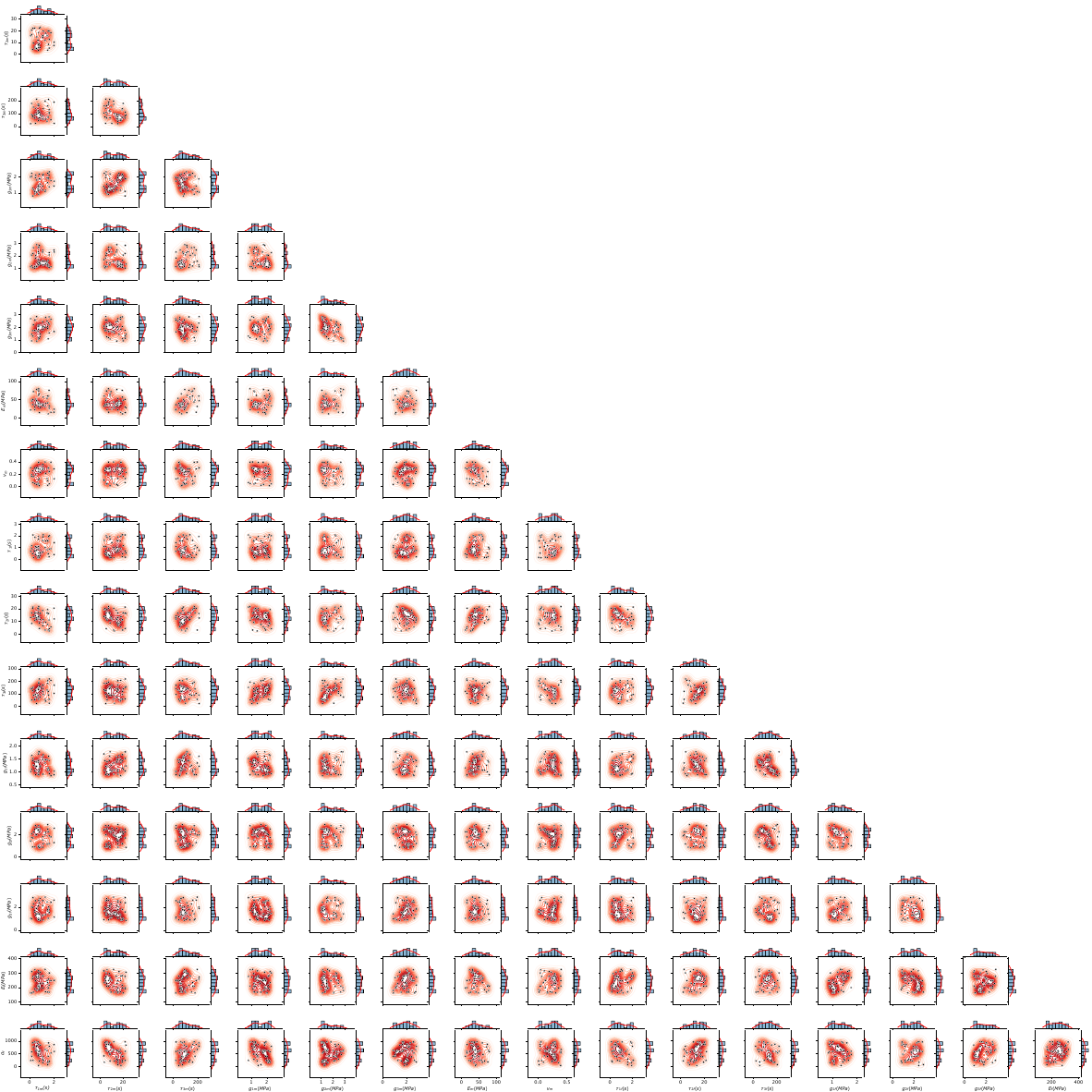}
\renewcommand{\thefigure}{S5} 
\caption {Posterior joint plots between parameters of the matrix and fiber component for the proximal site}
\label{Figure_S5}
\end{figure*}

\begin{figure*}[!ht]
\centering
\includegraphics [width=6in] {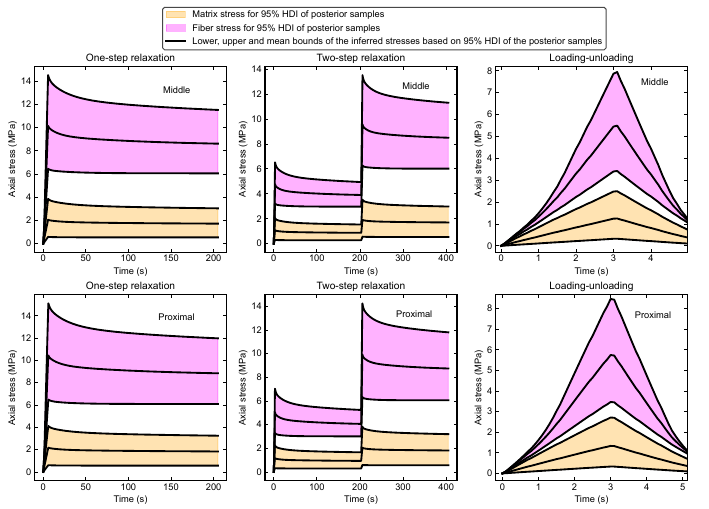}
\renewcommand{\thefigure}{S6} 
\caption {Inferred matrix and fiber stresses for the middle and proximal sites}
\label{Figure_S6}
\end{figure*}
\end{document}